%% file: main.tex
\documentclass[journal=tches, final]{iacrtrans}

\usepackage{graphicx}
\usepackage{amsmath,amssymb,amsfonts}
\usepackage{algorithmic}
\usepackage{textcomp}
\usepackage{xcolor}
\usepackage[utf8]{inputenc}
\usepackage{tikz}
\usepackage{pgfplots}
\usepackage{upgreek}
\usepackage{listings}
\usepackage{multirow}
\usepackage[acronym]{glossaries}
\usepackage{url}
\usepackage{subcaption}
\usepackage[hidelinks,bookmarks=false]{hyperref}
\usepackage[nohyperlinks, printonlyused]{acronym}
\usepackage{siunitx} 
\usepackage{dblfloatfix}
\usepackage{tikz}
\usepackage{makecell}
\usepackage{pgfplots}

\newcommand{\orcid}[1]{\href{https://orcid.org/#1}{\textcolor[HTML]{A6CE39}{\includegraphics[width=10pt]{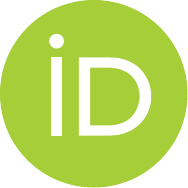}}}}

\usepackage{environ}
\makeatletter
\newsavebox{\measure@tikzpicture}
\NewEnviron{scaletikzpicturetowidth}[1]{%
  \def\tikz@width{#1}%
  \begin{lrbox}{\measure@tikzpicture}%
  \BODY
  \end{lrbox}%
  \pgfmathparse{#1/\wd\measure@tikzpicture}%
  \BODY
}
\makeatother

\author{Kathrin Garb\orcid{0000-0003-4139-5793} \and
Marvin Xhemrishi\orcid{0000-0001-5341-9954} \and
Ludwig Kürzinger\orcid{0000-0001-5312-3870}\and
Christoph Frisch
\authorrunning{K. Garb et al.}
\institute{Technical University of Munich, Munich, Germany \\
\email{kathrin.garb@tum.de} \\
\email{marvin.xhemrishi@tum.de} \\
\email{ludwig.kuerzinger@tum.de} \\
\email{chris.frisch@tum.de} }
}

\title{The Wiretap Channel for Capacitive PUF-Based Security Enclosures}
\titlerunning{The Wiretap Channel for Capacitive PUF-Based Security Enclosures}
\input{acronyms}

\begin{document}
\maketitle              
%

\keywords{Physical Unclonable Function \and Capacitive Enclosure \and Error Correction \and Wiretap Channel \and Polar Codes \and Physical Layer Security}

\begin{abstract}
In order to protect devices from physical manipulations, protective security enclosures were developed. However, these \emph{battery-backed} solutions come with a reduced lifetime, and have to be actively and continuously monitored. 

In order to overcome these drawbacks, \emph{batteryless} capacitive enclosures based on Physical Unclonable Functions (PUFs) have been developed that generate a key-encryption-key (KEK) for decryption of the key chain. In order to reproduce the PUF-key reliably and to compensate the effect of noise and environmental influences, the key generation includes error correction codes. However, drilling attacks that aim at partially destroying the enclosure also alter the PUF-response and are subjected to the same error correction procedures. Correcting attack effects, however, is highly undesirable as it would destroy the security concept of the enclosure. In general, designing error correction codes such that they provide tamper-sensitivity to attacks, while still correcting noise and environmental effects is a challenging task.

We tackle this problem by first analyzing the behavior of the PUF-response under external influences and different post-processing parameters. From this, we derive a system model of the PUF-based enclosure, and construct a wiretap channel implementation from $q$-ary polar codes.
We verify the obtained error correction scheme in a Monte Carlo simulation and demonstrate that our wiretap channel implementation achieves a physical layer security of 100 bits for 306 bits of entropy for the PUF-secret. 

Through this, we further develop capacitive PUF-based security enclosures and bring them one step closer to their commercial deployment. 
\end{abstract}


\input{secs/intro}
\input{secs/relatedWork}
\input{secs/systemAnalysis}
\input{secs/wiretap}
\input{secs/conclusion}

\section*{Acknowledgment}

The authors would like to thank DSO National Laboratories, Singapore, for publication permission of the measured and processed PUF-responses, and Fraunhofer AISEC for supporting the data acquisition.
Thanks to Peihong Yuan for providing us with good insights about the implementation.
Thanks also to Michal Virgovič for plotting scripts, and to Georg Maringer, Hedongliang Liu, and Matthias Hiller for valuable discussions.  \\
This work was partly funded by the European Research Council (ERC) under the European Union’s Horizon 2020 research and innovation program (Grant Agreement No. 801434), and by the German Ministry of Education and Research in the project VE-FIDES under grant number 16ME0257. 

\bibliographystyle{alpha}
\bibliography{Bibliography}

\end{document}

%% file: acronyms.tex
\newacronym{puf}{PUF}{Physical Unclonable Function}
\newacronym{hsm}{HSM}{Hardware Security Module}
\newacronym{pcb}{PCB}{Printed Circuit Board}
\newacronym{ic}{IC}{integrated circuit}
\newacronym{asic}{ASIC}{Application-Specific Integrated Circuit}
\newacronym{led}{LED}{Light-Emitting Diode}
\newacronym{ekms}{EKMS}{Embedded Key Management System}
\newacronym{trng}{TRNG}{True Random Number Generator}
\newacronym{lsb}{LSB}{Least Significant Bit}
\newacronym{kek}{KEK}{Key-Encryption-Key}
\newacronym{csp}{CSP}{Critical Security Parameter}
\newacronym{lmc}{LMC}{Limited Magnitude Code}
\newacronym{rs}{RS}{Reed Solomon}
\newacronym{nvm}{NVM}{Non-Volatile Memory}
\newacronym{cu}{Cu}{copper}
\newacronym{pi}{PI}{polyimide}
\newacronym{adc}{ADC}{Analog-to-Digital Converter}

\newacronym{vt}{VT}{Varshamov-Tenengolts}
\newacronym{bsc}{BSC}{Binary Symmetric Channel}
\newacronym{mse}{MSE}{Mean Squared Error}
\newacronym{sllc}{SLLC}{Systematic Low Leakage Coding}
\newacronym{ber}{BER}{Bit Error Ratio}
\newacronym{llr}{LLR}{log-likelihood ratio}
\newacronym{scl}{SCL}{successive cancellation list}
\newacronym{bec}{BEC}{Binary Erasure Channel}
\newacronym{bch}{BCH}{Bose Chaudhuri Hocquenghem}
\newacronym{ldpc}{LDPC}{Low-Density Parity-Check}
\newacronym{dmc}{DMC}{discrete memoryless channel}
\newacronym{scd}{SC}{successive cancellation}
\newacronym{hda}{HDA}{Helper Data Algorithm}
\newacronym{ser}{SER}{Symbol Error Rate}
\newacronym{fer}{FER}{Frame Error Rate}
\newacronym{fib}{FIB}{Focused Ion Beam}
\newacronym{lfi}{LFI}{Laser Fault Injection}

%% file: secs/intro.tex
\section{Introduction}

Devices deployed for high-security applications, such as \glspl{hsm} \cite{NIST-SP800-57}, have to withstand tampering attacks. For this purpose, cryptographic modules 
require a physical boundary in the form of enclosures or coatings \cite{Fed08, ISO-24759, ISO-19790}
that can detect a tamper event and trigger an alarm, zeroing all relevant \glspl{csp} on the device.

The first generation of enclosures contained meshes of electrodes that continuously monitor resistivity to determine whether a tamper-event occurred \cite{OI18, Gore-cover, Gore-envelope, FIPS-IBM}.
However, monitoring required a continuous power supply in the form of batteries, which reduced the overall lifetime of the device and came with a higher sensitivity towards environmental influences.

To overcome these drawbacks, capacitive meshes based on \glspl{puf} have been developed without the need for a continuous power supply \cite{B-TREPID, COVER}.
A \gls{puf} is an object's fingerprint that stems from manufacturing variations \cite{MaesPhDThesis}, for instance, of electrodes that form the capacitive mesh within the enclosure. 
The capacitive differences within the mesh constitute the PUF-response, from which a \gls{kek} is generated to decrypt further \glspl{csp} of the key hierarchy \cite{EKMS, FORTRESS}.
Tampering attacks damage the enclosure and change electrode capacitance, which will destroy the PUF-key, and, thus, lead to a loss of all \glspl{csp}. 
However, also environmental changes affect the measured PUF-response and have to be compensated to a certain degree through error-correction codes. 

This ``dilemma'' of ensuring tamper-sensitivity to attacks, while allowing environmental changes within a certain range is challenging.
Not only does it require the integration of suitable error correction codes into the \gls{puf} post-processing, but what is more, a mechanism to model both cases separately. 
We address this problem through the following contributions. \newline

\textbf{Our contributions:} 
\begin{itemize}
    \item We characterize the effects of temperature changes and tampering attacks on the PUF-based capacitive enclosure. Based on the obtained data, we discuss distortion and the impact of an attack on the PUF-based capacitive enclosure using appropriate quantization schemes.
    \item We introduce q-ary polar codes for error-correction of our Higher Order Alphabet PUF and apply a wiretap code on the practical data, obtained from measurements of the PUF-response, to protect the PUF-secret from an attacker. We also discuss the differences between our code and other wiretap codes for \glspl{puf}.
    \item We demonstrate how to achieve a physical-layer security level of up to 100 bits, for a PUF-secret of 306 bit length.    
\end{itemize}

\textbf{Overview:} \newline
In Section~\ref{sec:relatedWork}, we give an overview of tamper-sensitive \glspl{puf}, and relevant error correction codes.
This is followed by an analysis of the PUF-response influenced by temperature changes and drilling attacks in Section~\ref{sec:keyGenEnclosure}.
We include the effects of different choices of quantization, determining suitable parameters that model the capacitive enclosure.
We also discuss different key generation schemes and their suitability in the context of the capacitive PUF-based security enclosure.

Section~\ref{sec:wiretapChannel} proposes a solution to the described dilemma.
Based on our system analysis in Section~\ref{sec:keyGenEnclosure}, we construct a wiretap channel through $q$-ary polar codes and verify our results in a Monte Carlo simulation.
We conclude our work in Section~\ref{sec:conclusion}.

%% file: secs/relatedWork.tex
\section{Related Work}
\label{sec:relatedWork}

This section discusses tamper-sensitive \glspl{puf} and relevant error correction codes. 
We also highlight state-of-the-art approaches, where wiretap coding previously has been used for PUFs to improve the overall security.

\subsection{Tamper-Sensitive Physical Unclonable Functions}

The minuscule manufacturing variations of the \gls{puf} and physical unclonability make it suitable for tamper-sensitive applications. 

One of the first larger scale \gls{puf}-based secure architectures is the optical waveguide polymer \gls{puf} \cite{OpticalPUFSystem,OpticalPUF}, where a polymer waveguide covers the top area of a \gls{pcb}.
To detect whether a tamper event occurred, light from \glspl{led} is sent through the polymer waveguide, and the created light patterns are analyzed.
A drawback of the optical waveguide polymer \gls{puf} is, however, that it does not protect the whole device but only the top area of the \gls{pcb}.
Optical \glspl{puf} were also applied to smartcards by Esbach et al. \cite{EFK+12}.

Another electromagnetic \gls{puf} arranges antennas within an electromagnetic-sensitive sealing material \cite{TZP20}.
The wavelength of the radio signals varies
due to manufacturing variations of the sealing material.
The measurement of the channel state information between antennas determines whether a tamper event occurred.
A similar approach is the anti-tamper radio proposed by Staat et al. \cite{Staat2021}.

Smaller-scale \glspl{puf} coatings aim to protect small areas on \glspl{ic}.
One such example was proposed by Tuyls et al.~\cite{CoatingPUF}, where the capacitance of randomly arranged particles within a coating is measured in order to determine a tamper event.  

Zhang et al. \cite{ZHW+21} recently proposed Switched-Capacitor \glspl{puf} as protection of electrical circuits from manipulation.

Another technology closely related to challenge-response \glspl{puf} is a 3D hardware canary introduced by Briais et al.\ \cite{Briais2012}. Here, a security-sensitive circuit is surrounded by a wire cage. A spatially distributed chain of functions placed at the vertices of the cage forms the hardware canary. A challenge has to be answered by a correct response to attest to the canary's integrity. The main difference between \glspl{puf} and hardware canaries is that hardware canaries are constructed through an algorithm, while \glspl{puf} are based on random manufacturing variations.

Capacitive PUF-based enclosures \cite{B-TREPID,COVER} protect larger areas and entire \glspl{pcb}, as discussed in more detail in Section~\ref{sec:keyGenEnclosure}.

\subsection{Error Correction Codes}

The application of Physical Unclonable Functions requires a sufficient level of reliability under environmental changes.
For this purpose, PUF-key generation includes an error correction step.
There are various error correction codes of \glspl{puf} that focus on optimizing the code in terms of implementation overhead, decoding complexity and hardware resources \cite{BGS+08,MVHV12,HKS20,MHK+19,PMB+15} applying \gls{bch}, Reed Solomon or Reed Muller Codes.
Another way of reducing the implementation overhead is by including soft and reliability information \cite{MTV09a, MTV09b, MPB18, HOSB16}. 

Apart from implementation efficiency, secrecy leakage is a known issue for \glspl{puf}.
Hence, several schemes were proposed to tackle this problem, focusing either on information leakage due to bias in the PUF-response \cite{MLSW16, BY19, IHL+19} or on the reduction of helper data leakage \cite{CW19, Bai2021}. 
Müelich and Bossert circumvented the leakage problem by proposing a novel secure sketch, where no additional helper data is required \cite{MB17}.
One of the main goals of the error correction code is to optimize the failure probability.
This was the objective of Chen et al., who applied Polar Codes to SRAM \glspl{puf} \cite{CIW+17}.

In the context of PUF-based enclosures, achieving tamper-sensitivity is a significant issue, as we will see in the subsequent section.
An attempt to ensure high-sensitivity to attacks while still correcting environmental effects was proposed by Immler and Uppund, who applied \glspl{lmc} \cite{LMC}, where only errors of a certain magnitude are corrected. 

\subsection{Wiretap Coding and PUFs}
The task of an error correction code for \glspl{puf} is to target the noise effects in the PUF-response to derive an error-free secret key. Wiretap codes, on the other hand, have an additional feature: not only can they correct errors in the noisy \gls{puf}-response, but they also incorporate a security aspect.
Wyner has shown in \cite{WynerI} that additional randomness besides the error correction capability is necessary to achieve security. Consequently, not every error correction code applied to \glspl{puf} is a suitable wiretap code. Hence, several papers have studied wiretap codes, in particular for \glspl{puf}.
They have been introduced for PUFs in \cite{hiller2017hiding} for codes of up to length 64 bit.
In \cite{BY19} and \cite{Bai2021} the authors extend this preceding work to larger binary polar codes.
However, in comparison to the proposed approach in this work, there are three major differences:
(i) In the state-of-the-art applications of wiretap coding for PUFs, the goal is debiasing instead of physical layer security.
(ii) The attacker in the state-of-the-art approaches is assumed to be weaker than in this work:
In \cite{hiller2017hiding, BY19, Bai2021}, the attacker receives the helper data as output of his channel.
Hence the whole PUF-response is interpreted as error induced over the wiretapper's channel.
In this work, the received message is degraded in comparison to the legitimate channel only based on the impact of a drilling attack.
(iii) We operate on $q$-ary instead of binary polar codes; the enclosure \gls{puf} in this work allows for a higher-order quantization, and thus, more entropy can be extracted from the \gls{puf} than in a binary case.
Consequently, we propose a novel scheme because any \gls{puf} error correction cannot detect an attacker, and the previous wiretap codes for \glspl{puf} do not match our attacker model and the non-binary response of our enclosure \gls{puf}. This is detailed in the following sections.

%% file: secs/systemAnalysis.tex
\section{Key Generation for Capacitive PUF-Based Enclosures}
\label{sec:keyGenEnclosure}

The capacitive PUF-based enclosure requires several blocks: The measurement of the PUF-response (\ref{subsec:capacitiveEnclosure}), processing of the analog PUF-data before quantization (\ref{subsec:analysisPUF-response}), quantizing analog data to symbols (\ref{subsec:quantization}), and finally, error-correction on these symbols including key generation (\ref{subsec:keyGen}).

Previous work on the capacitive enclosure, including an overview of the system components, is summarized in Subsection~\ref{subsec:capacitiveEnclosure}. An overview of post-processing steps, as proposed in \cite{COVER, B-TREPID}, is given in~\ref{subsubsec:generalAnalysis}. 

To enhance the existing PUF-model, we provide a detailed analysis of the impact of temperature effects and drilling attacks on the enclosure in Section~\ref{subsec:analysisPUF-response}. Our analysis is based on the measured PUF-responses obtained from~\cite{COVER}. In Section~\ref{subsec:quantization}, we analyze the distortion and error for different types of quantization. The goal of our analysis is to obtain an estimate for the error probability of our channel model and to investigate binary and $q$-ary channel models. Section~\ref{subsec:keyGen} focuses on the different possible key generation schemes for the PUF-model and $q$-ary channels.

\subsection{The Capacitive Enclosure}
\label{subsec:capacitiveEnclosure}

\begin{figure}[t!]
\centering
\includegraphics[width=0.55\textwidth]{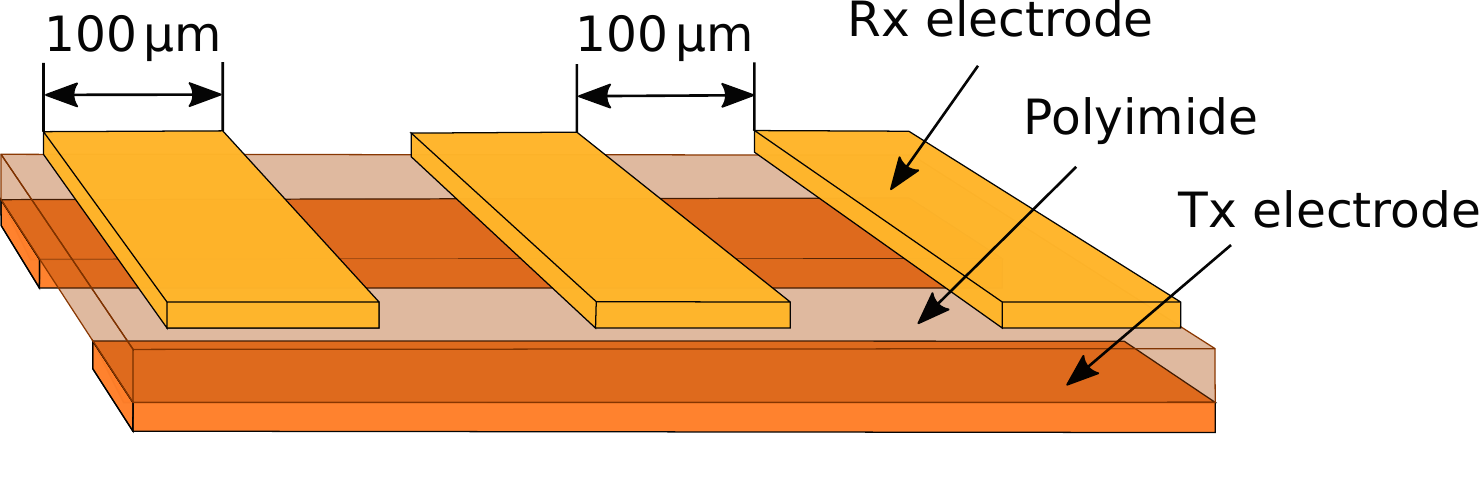}
\caption{Schematic cross section of the envelope mesh with overlapping Rx and Tx electrodes.}
\label{fig:envelopeMesh}
\end{figure}

\begin{figure}[b!]
\centering
\includegraphics[width=0.65\textwidth]{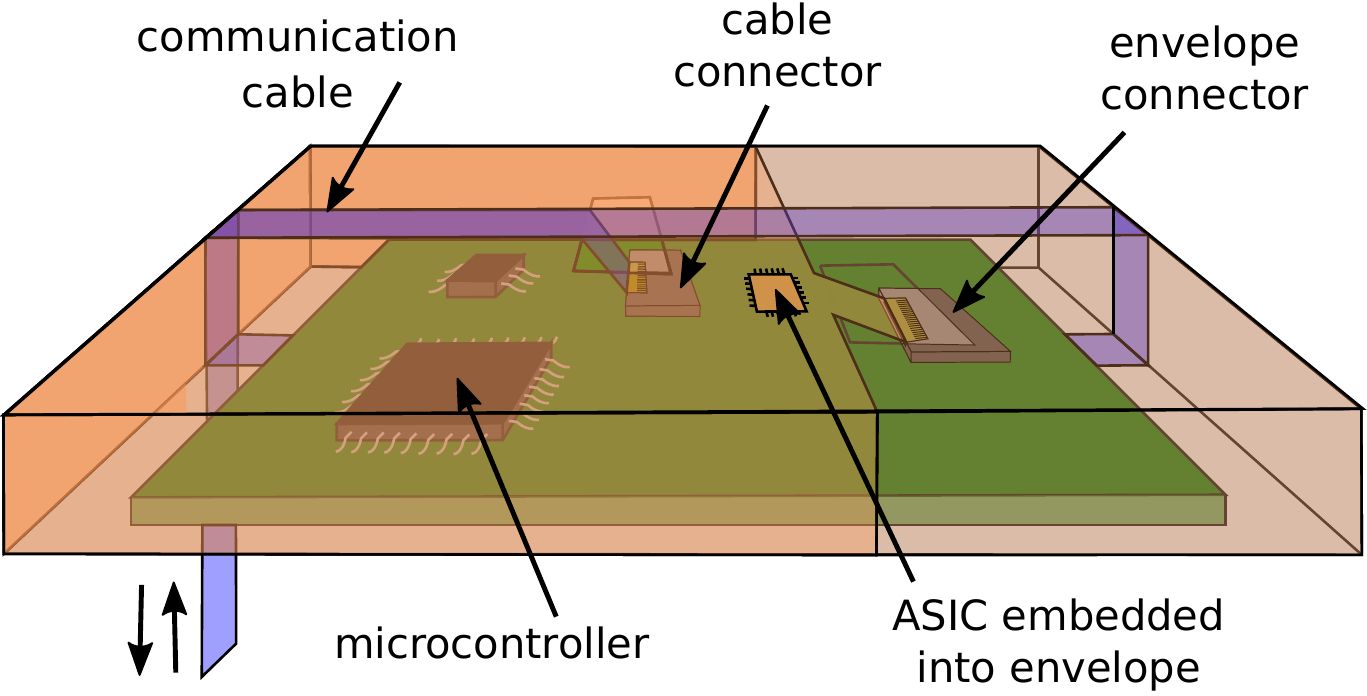}
\caption{Schematic overview of the capacitive envelope protecting a PCB.}
    \label{fig:envelope}
\end{figure}

The $\SI{18.5}{\centi\meter}\times\SI{9}{\centi\meter}$ capacitive envelope (B-TREPID) \cite{B-TREPID} consists of two layers, each with 16 \gls{cu} electrodes (Rx and Tx) --- arranged in a meander structure --- that are separated by an insulating layer of \gls{pi}.
The electrodes have a width and distance of \SI{100}{\micro\meter}, as depicted in Figure~\ref{fig:envelopeMesh}.
To reduce the impact of alternating electric fields on the capacitive measurement, the top and bottom of the envelope were extended by an additional shielding layer of \gls{cu} \cite{ObermaierPhD, B-TREPID}, leading to an overall thickness of approximately \SI{0.3}{\milli\meter}.
Together with a cable for external communication, the envelope is wrapped around a casing containing the protected \gls{pcb}, depicted in Figure~\ref{fig:envelope}. 

The \gls{cu} electrodes overlap and form $16 \times 16$ \emph{absolute} capacitances, which are, however, not suitable for the evaluation as a \gls{puf} since they depend on global manufacturing variations. 
The actual PUF-response is, hence, formed by measuring the difference between these absolute values, resulting in 128 \emph{differential} capacitances with an estimated maximum entropy of 560 bits.
These differential capacitances are obtained by measuring two Tx-electrodes (Tx pair) against one Rx-electrode.
From the PUF-response, a secret \gls{kek} is derived (key enrollment) and repeatedly reproduced (key reproduction).
A drilling attack with a diameter of \SI{300}{\micro\meter} destroyes two of the electrodes, leading to a loss of 80 bits of entropy~\cite{B-TREPID} in the unprocessed PUF-response, which alters the reproduced key. 

To measure the PUF-response, a discrete measurement circuit was developed \cite{measurementCircuit}.
Since this circuit was optimized in terms of accuracy and not size, a smaller measurement \gls{ic} of approximately $\SI{5}{\milli\meter}\times\SI{5}{\milli\meter}$ was developed and integrated into the envelope \cite{ASIC}.
The \gls{ic} measures 16 differential capacitances in parallel, which decreases the measurement time.
However, the optimization in terms of space and time comes at the expense of accuracy.

A microcontroller carries out the post-processing of the PUF-response on the \gls{pcb} running the \gls{ekms} \cite{EKMS, FORTRESS}.
The \gls{ekms} is a hardened FreeRTOS with integrated key management and key generation.
This also includes the error correction algorithm to compensate for noise and environmental effects.

To cover only parts of a device, an enclosure (COVER) was developed, which differs from the envelope (B-TREPID) in its use case and mesh arrangement. 
COVER, in contrast to the envelope, is not wrapped around the device but only covers the top or bottom area of the \gls{pcb}. 

We measured and statistically analyzed the PUF-distribution of 50 envelopes (B-TREPID) \cite{B-TREPID} and 115 enclosures (COVER) \cite{COVER}. 
The differential capacitances of B-TREPID and COVER are both Gaussian distributed and behave equivalently under external influences. However, their overall range in femtofarads differs, which makes merging both data sets difficult. In general, the distribution of COVER is broader compared to B-TREPID.
Since the COVER data is more extensive than the B-TREPID data, the following analysis will focus on data obtained from the COVER.

\subsection{Analysis of the PUF-Response}
\label{subsec:analysisPUF-response}

This section covers an analysis of the measured capacitances and how they are affected by temperature changes and drilling attacks.

\subsubsection{General Overview}
\label{subsubsec:generalAnalysis}

For the following analysis, the PUF-response was measured with the discrete circuit \cite{measurementCircuit} instead of the \gls{ic} \cite{ASIC} to increase accuracy.
Furthermore, since more data is available for COVER \cite{COVER} compared to the B-TREPID \cite{B-TREPID} and since both systems behave equivalently, the PUF-responses for the following analysis stem from COVER.

The 128 unprocessed differential capacitances are integers in the interval $\left[-10000, +10000\right]$ that are Gaussian distributed with a standard deviation of $\sigma = \SI{30}{\femto\farad}$ (corresponding to 2241 points) after normalization (see Post-Processing) measured over 115 COVERs \cite{COVER}.  
The $\left[-10000, +10000\right]$ interval corresponds to a full-scale range of $\left[-\SI{134}{\femto\farad}, +\SI{134}{\femto\farad}\right]$ with a digital resolution of \SI{13.4}{\atto\farad} corresponding to 1 point \cite{ObermaierPhD}. 
Measuring a single differential capacitance --- also called ``node'' --- with the discrete circuit takes \SI{390}{\micro\second} on average, while the differential measurement time for the entire enclosure amounts to \SI{50}{\milli\second}. \newline

\input{secs/post-processing}

\textbf{Noise Distribution}\newline
Despite low-noise components and noise filtering within the measurement circuit, an inevitable noise caused by the full setup of the enclosure and the measurement system still remains. 
The noise plays a major role in the quantization of the PUF-response, which is discussed in Subsection~\ref{subsec:quantization}.
The noise distribution was statistically determined from 200 consecutive measurements of the same COVER.
It is well-approximated through a Gaussian with a standard deviation of \SI{1.7}{\femto\farad} corresponding to 129 points \cite{ObermaierPhD} centered around $\mu = \SI{0}{\femto\farad}$.
\newline

\textbf{Post-Processing}\newline
After measuring the PUF-response, several post-processing steps are necessary before the generation (or reproduction) of the PUF-key.
Figure~\ref{fig:post-processing} depicts the three steps of post-processing with analog helper data generation:
\begin{itemize}
\item[1.] Shift of TX groups (normalization)
\item[2.] Generation of quantization mapping
\item[3.] Generation of analog helper data 
\end{itemize}

The raw differential capacitances are obtained by measuring two Tx electrodes against one Rx electrode.
We will refer to each of these Tx-pairs measured against all other Rx electrodes as a Tx group.  
However, \emph{global} manufacturing variations in the thickness of the electrodes lead to offsets in the raw differential capacitance of some of the Tx groups \cite{measurementCircuit}.
In order to remove the dependency on these global variations of each Tx group, we subtract the group mean in the first step. We will refer to the Tx group shift as ``normalization''.

In the second step, the PUF-response is quantized, hence, it is divided into quantization intervals. 
To reduce the quantization error, the PUF-values of each interval are shifted to the center of the interval (step 3), which further reduces the quantization error (see \cite{IHKS16}).
This offset is stored as \emph{analog helper data}. 
Even though, this step reduces the quantization error, it is not sufficient to compensate all errors stemming from noise, let alone environmental changes, as we will see in the following.
Subsection~\ref{subsec:quantization} will discuss quantization in more detail.

\subsubsection{Temperature Distribution}

\begin{figure}[b!]
\centering
\includegraphics[width=0.6\textwidth]{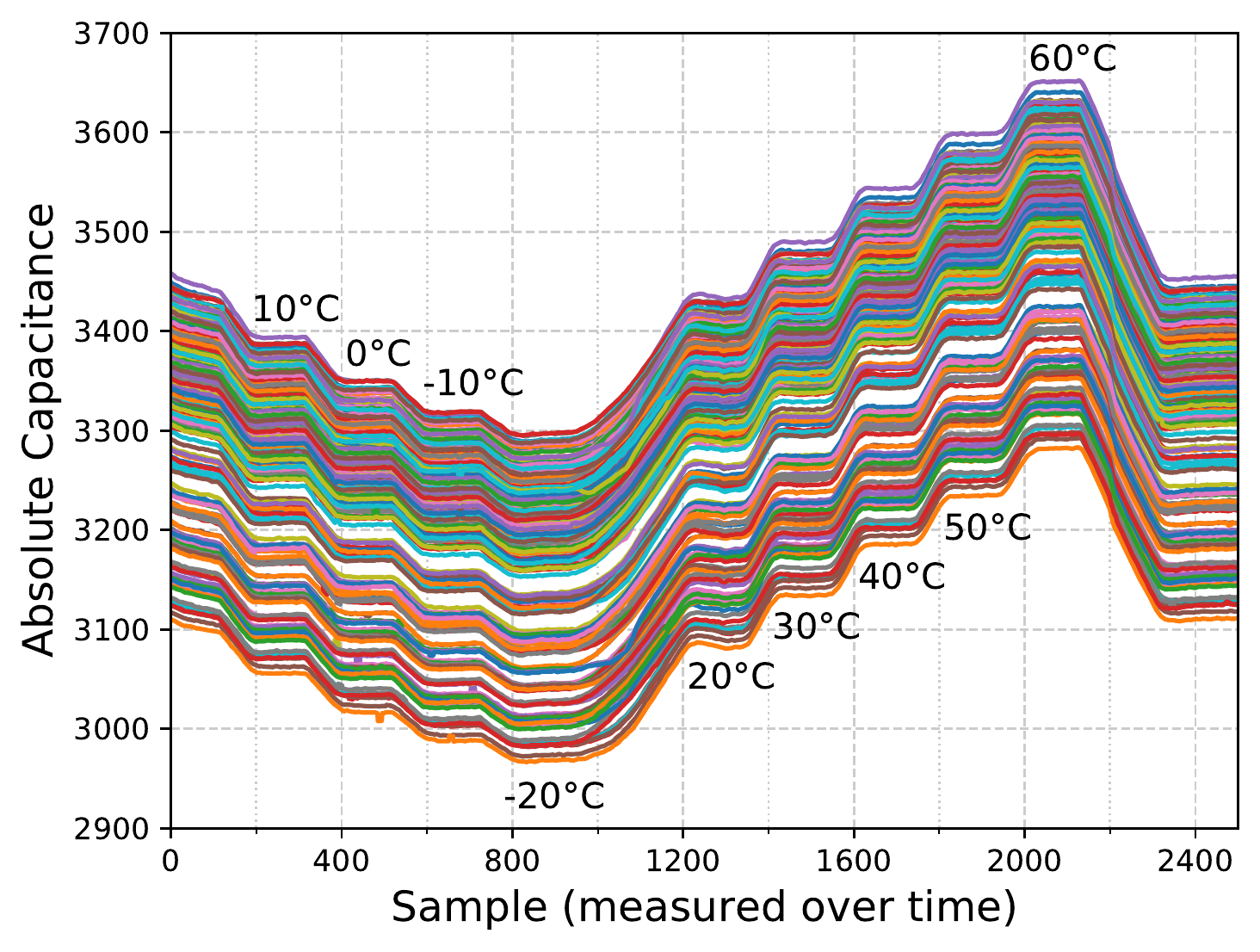}
\caption{Absolute capacitance under temperature change. Data taken from \cite{COVER}.}
\label{fig:absTemp}
\end{figure}

\begin{figure}[t!]
\centering
\hspace*{-0.2cm}
\begin{subfigure}{0.5\textwidth}
    \centering
    \includegraphics[width=\textwidth]{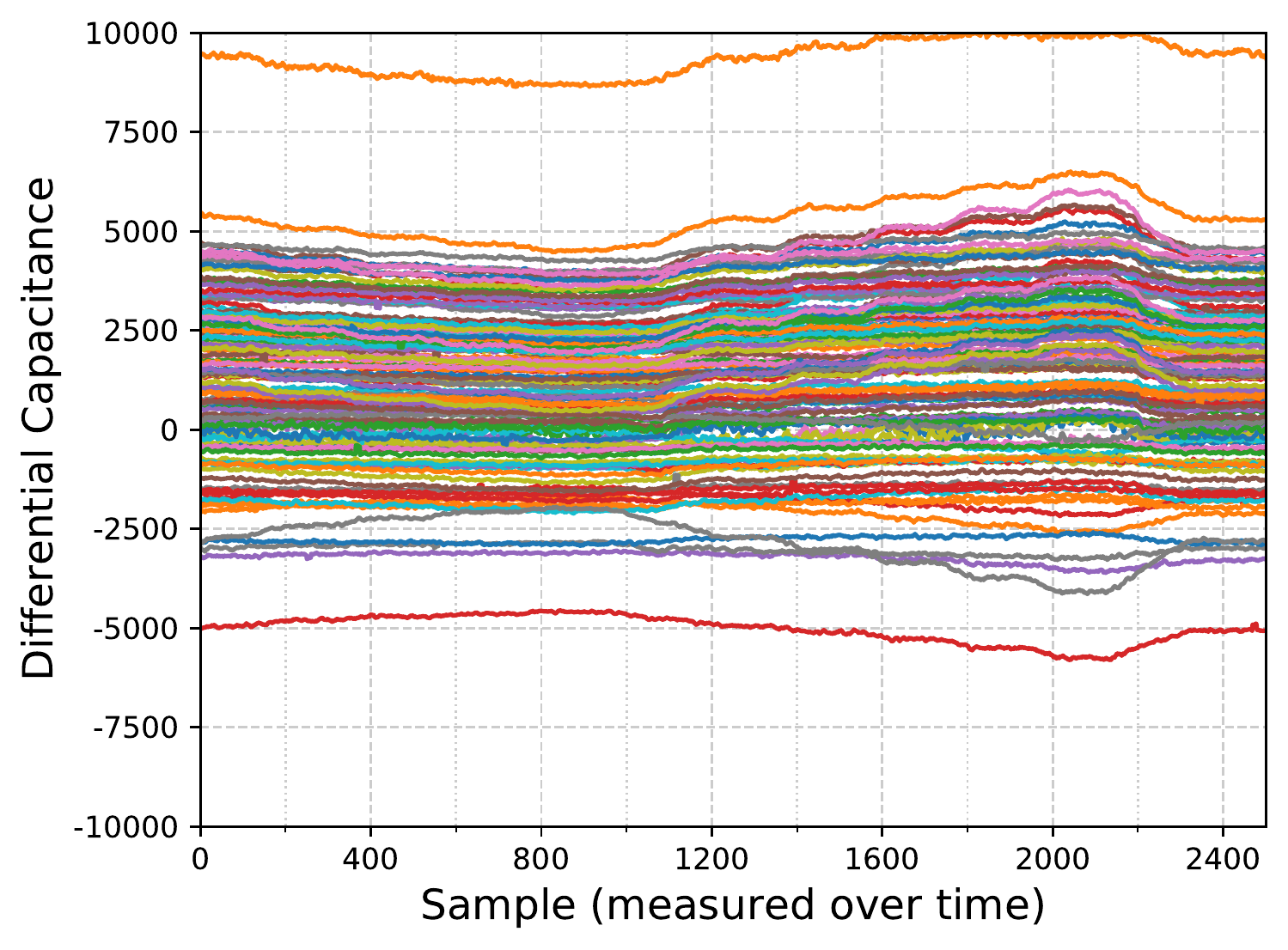}
    \caption{Raw differential capacitance under temperature change (compare to Fig.~\ref{fig:absTemp}). Data taken from \cite{COVER}.}
    \label{fig:tempRawDiff}
\end{subfigure}\hspace*{0.3cm}%
\begin{subfigure}{.5\textwidth}
    \centering
    \includegraphics[width=\textwidth]{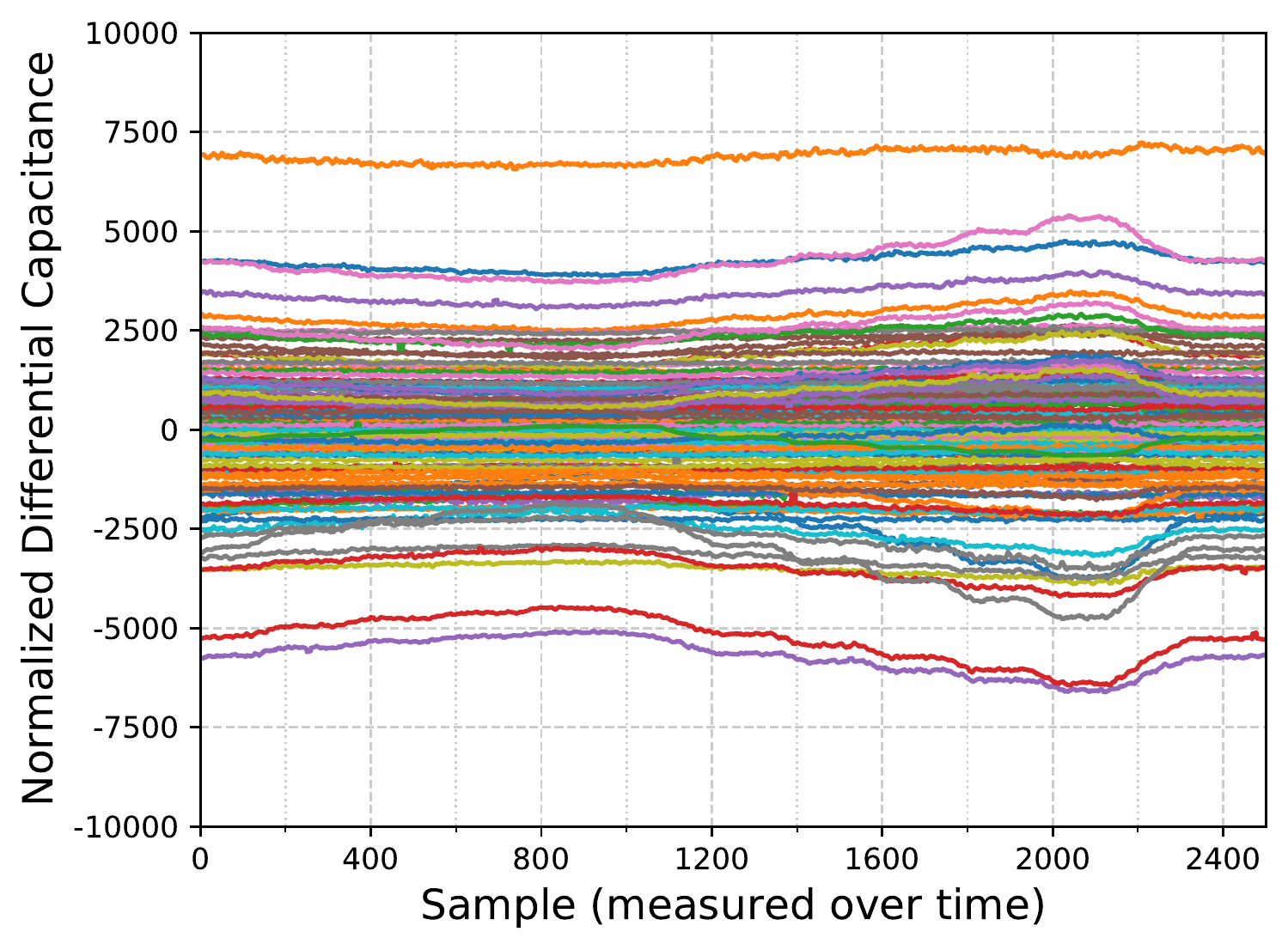}
    \caption{Differential capacitance under temperature change \emph{after} Tx group shift. \newline $\phantom{.}$}
    \label{fig:tempRawDiffShift}
\end{subfigure}
\hspace*{-0.1cm}\begin{subfigure}{.5\textwidth}
    \centering
    \includegraphics[width=\textwidth]{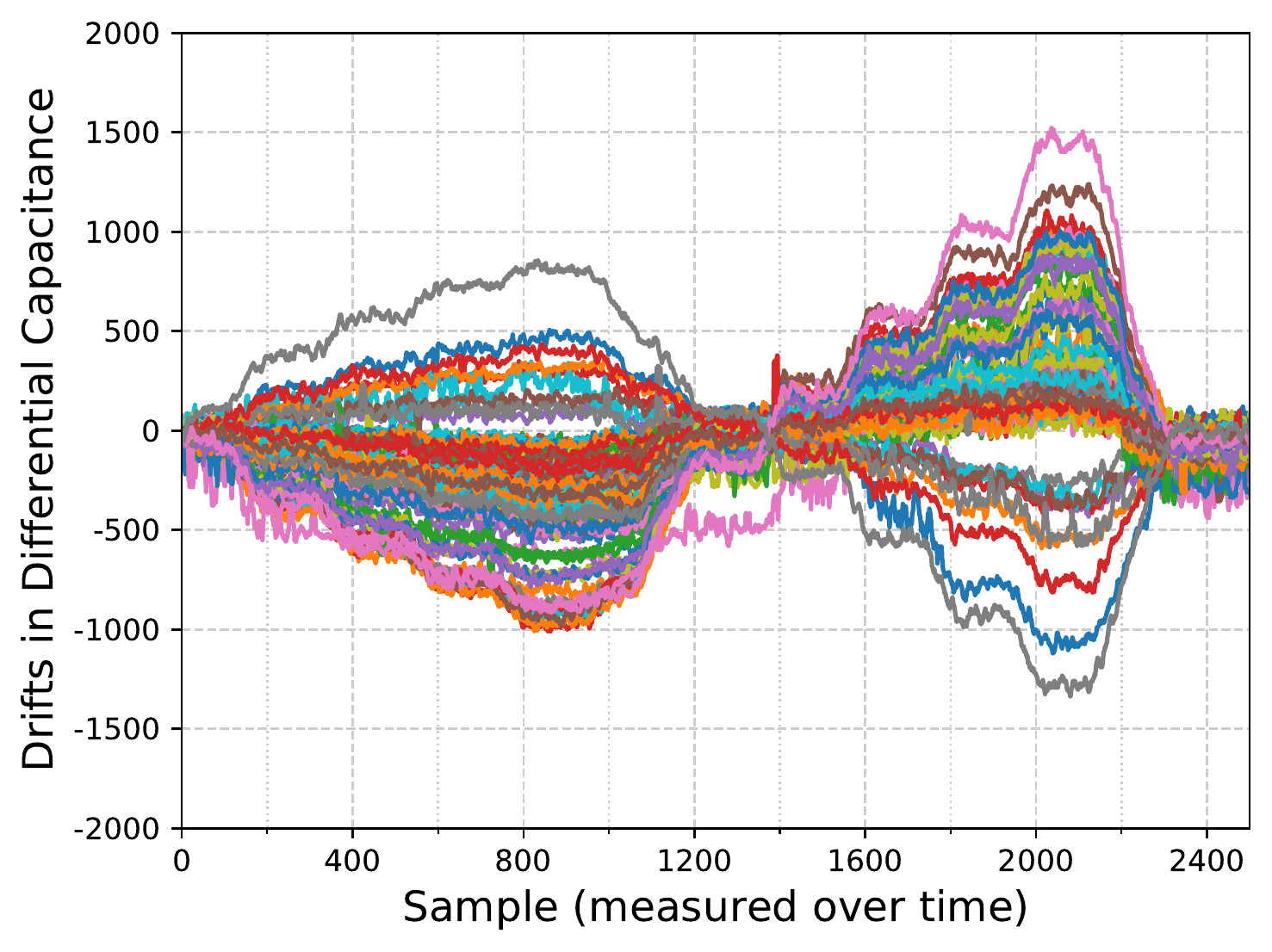}
    \caption{Change in differential capacitance due to temperature changes compared to reference measurement (at \SI{20}{\celsius}) \emph{before} Tx group shift.}
    \label{fig:tempDiffDrift}
\end{subfigure}\hspace*{0.3cm}%
\begin{subfigure}{.5\textwidth}
    \centering
    \includegraphics[width=\textwidth]{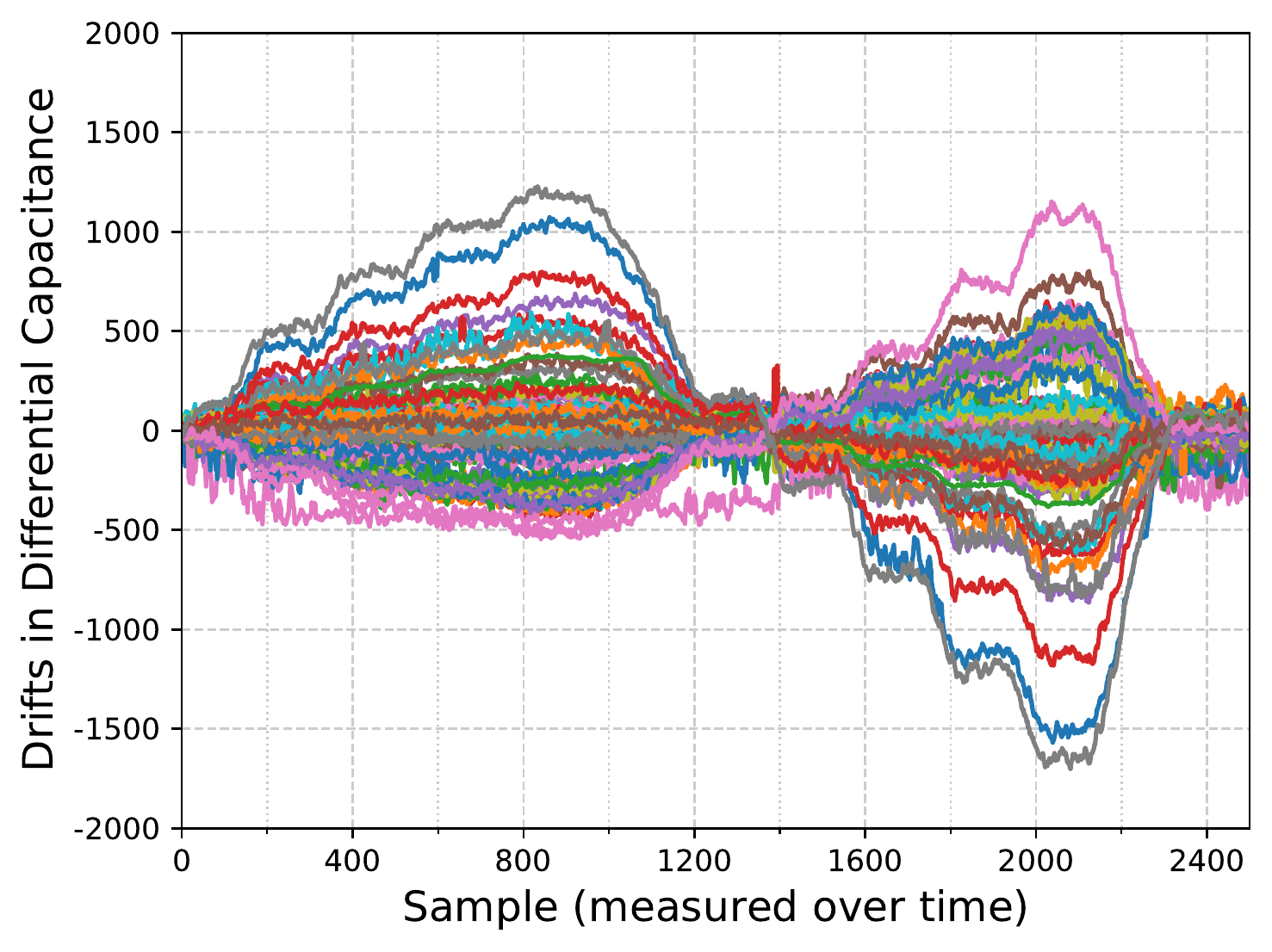}
    \caption{Change in differential capacitance due to temperature changes compared to reference measurement (at \SI{20}{\celsius}) \emph{after} Tx group shift.} 
    \label{fig:tempDiffDriftShift}
\end{subfigure}
\hspace*{-0.1cm}\begin{subfigure}{.5\textwidth}
    \centering
    \includegraphics[width=\textwidth]{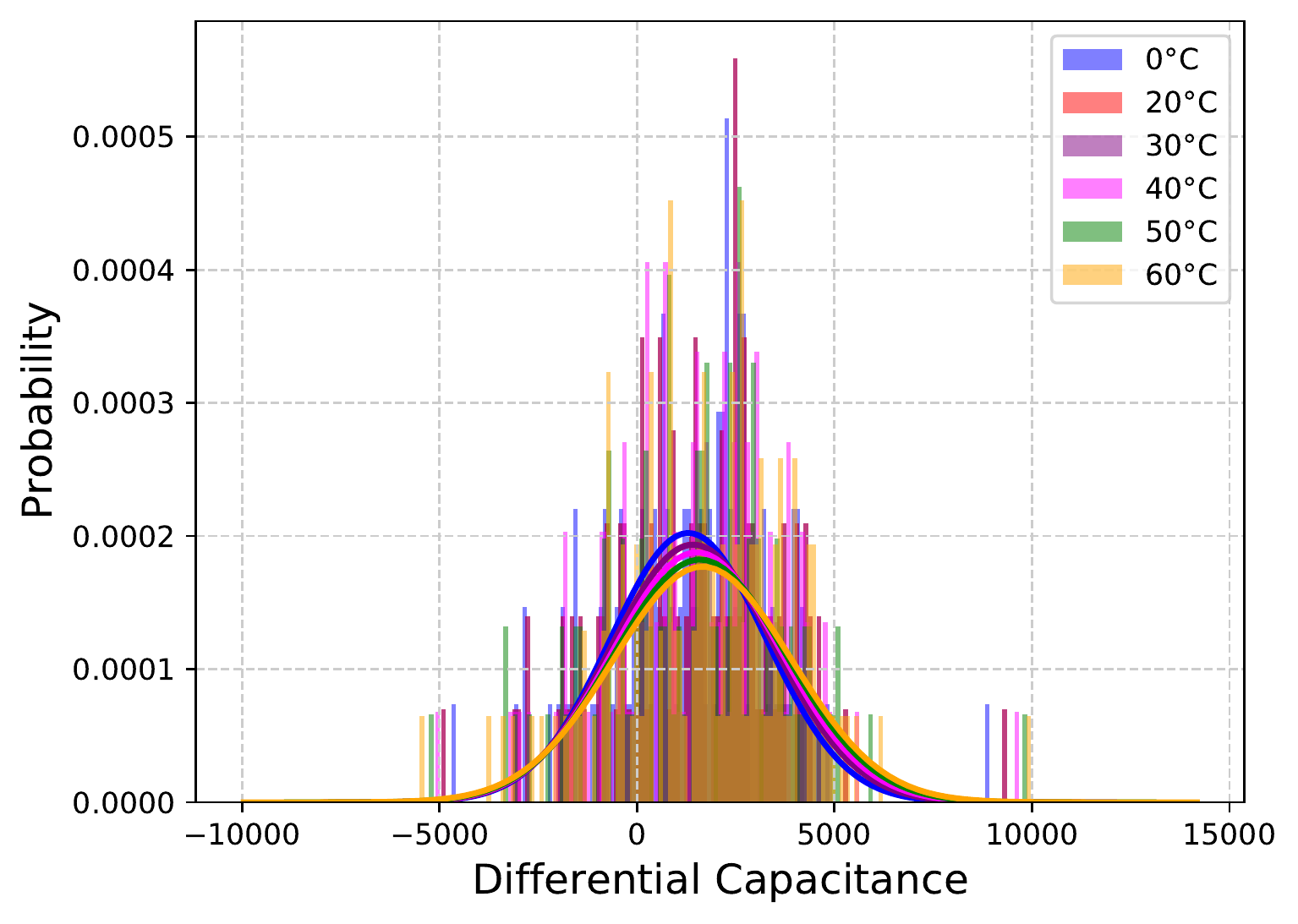}
    \caption{Distribution of the differential capacitance (with Gaussian fit) for temperatures between \SI{0}{\celsius} and \SI{60}{\celsius} \emph{before} Tx group shift.}
    \label{fig:temphist}
\end{subfigure}\hspace*{0.3cm}%
\begin{subfigure}{.5\textwidth}
    \centering
    \includegraphics[width=\textwidth]{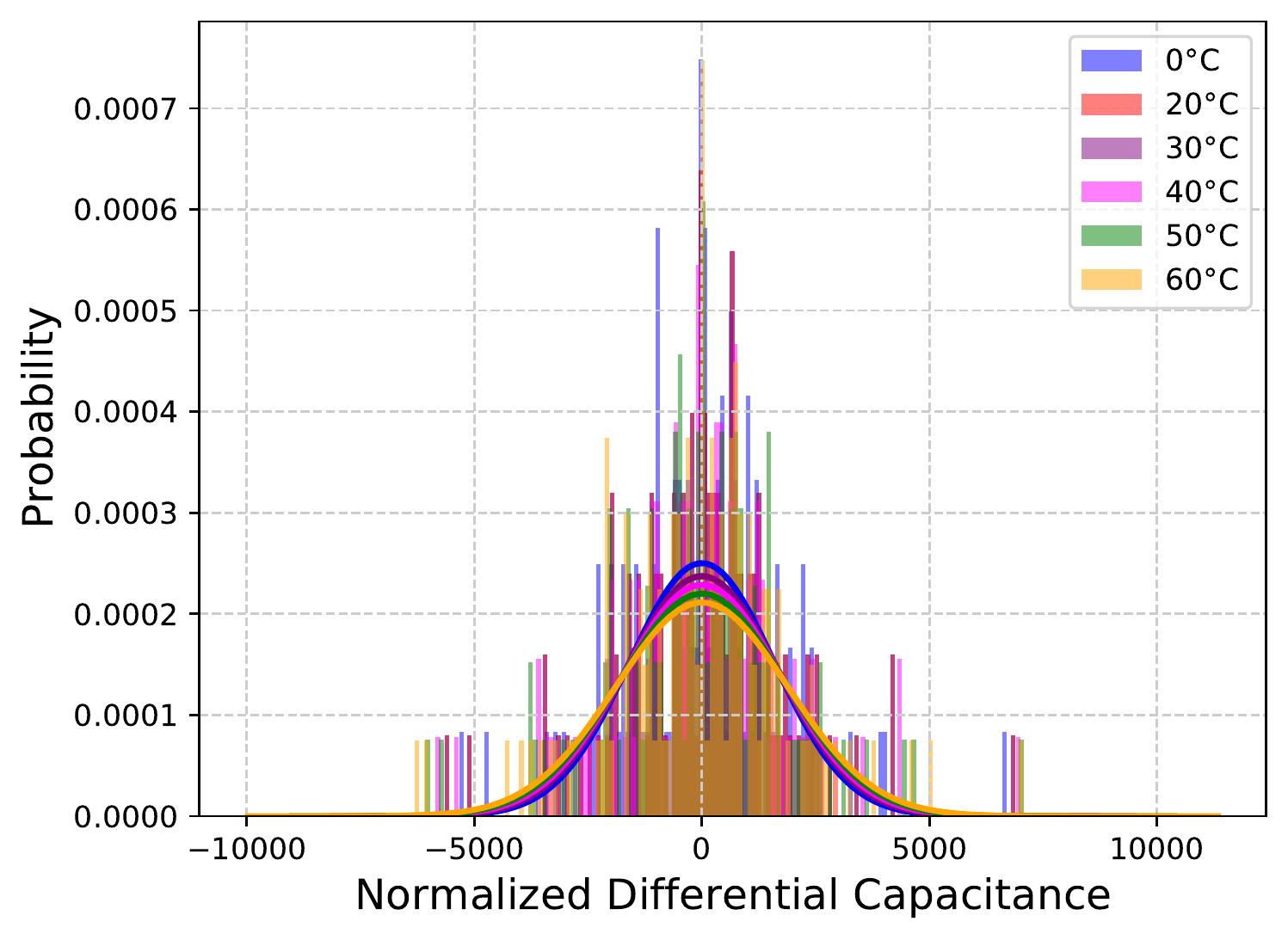}
    \caption{Distribution of the differential capacitance (with Gaussian fit) for temperatures between \SI{0}{\celsius} and \SI{60}{\celsius} \emph{after} Tx group shift.} 
    \label{fig:temphistDrift}
\end{subfigure}
\caption[short]{Analysis of the differential capacitance under temperature changes.}
\end{figure}

Environmental effects, for example, changes in temperature and humidity, alter the measured PUF-response. 
The effects of electromagnetic interference were considered in the design of the \gls{puf} and the measurement circuit. A copper shielding was added to the envelope to counteract the effect of alternating electric fields. Without the shielding, reliably measuring the differential capacitances is not possible. The effects of external magnetic fields are counteracted by the narrow excitation frequency and canceled out through the meander structure of the \gls{puf}. Hence, for the following analysis, we focus solely on temperature effects.
Fig.~\ref{fig:absTemp} shows the absolute capacitances for temperatures between \SI{-20}{\celsius} and \SI{+60}{\celsius}. 
The corresponding differential capacitances are depicted in Fig.~\ref{fig:tempRawDiff} before normalization (Tx group shift) and in Fig.~\ref{fig:tempRawDiffShift} after. 

Apart from a shifted mean, the normalized distribution (Fig.~\ref{fig:tempRawDiffShift}) also appears more narrow than the distribution of the raw capacitances (Fig.~\ref{fig:tempRawDiff}).
The normalized distribution exhibits a lower variance, as all Tx group means are shifted separately.

Fig.~\ref{fig:tempDiffDrift} depicts the change in the measured differential capacitance compared to a reference distribution at \SI{20}{\celsius}.
The maximum difference amounts to 1500 points at \SI{60}{\celsius} (Fig.~\ref{fig:tempDiffDriftShift}), where most node changes are within $\left[-700, +700\right]$ points. 
Fig.~\ref{fig:temphist} and Fig.~\ref{fig:temphistDrift} depict histograms of the differential capacitance for temperatures between \SI{0}{\celsius} and \SI{60}{\celsius} before and after normalization.
The Gaussian fits show that --- in both cases --- as the temperature increases, the distribution broadens, even though the standard deviation is of a smaller magnitude ($\Delta \sigma = 207$) after normalization. 
In the case of the raw distribution, the mean is shifted to the right as the temperature increases.

The enclosure protects devices that have to resist large environmental changes in the field that depend on the reliability requirements for the particular use case.
For instance, the security policy for the HP Atalla Ax160 PCI HSM \cite{atalla-temperature} states that for temperatures outside of the range $\left[ \SI{-20}{\celsius}, \SI{+100}{\celsius}\right]$, a tamper-event is generated. 
Hence, an envelope or COVER protecting such a device will have to withstand temperature changes outside of the interval $\left[\SI{-20}{\celsius}, \SI{+60}{\celsius}\right]$.
Larger temperatures create even larger offsets in the raw differential data and, thus, broaden the distribution.

\subsubsection{Attack Distribution}

\begin{figure}[t!]
    \centering
    \includegraphics[width=0.5\textwidth]{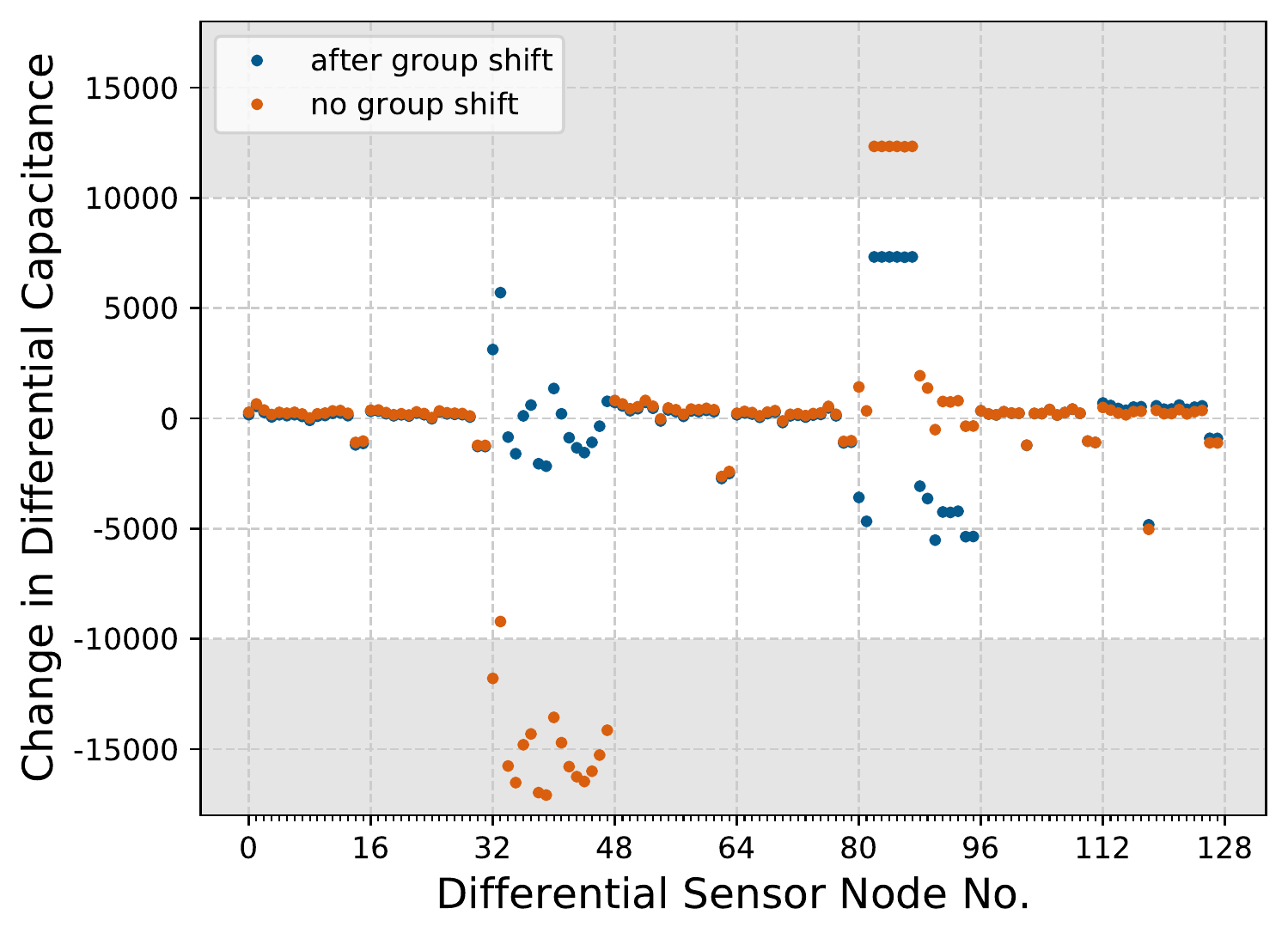}
    \caption{Change in differential capacitance for a drilling attack affecting two Tx groups (before and after normalization). 
The grey area denotes values within the range of non-linear effects.}
    \label{fig:attack_change_first}
\end{figure}

\begin{figure}[b!]
\centering
\hspace*{-0.1cm}\begin{subfigure}{0.5\textwidth}
    \centering
    \includegraphics[width=\textwidth]{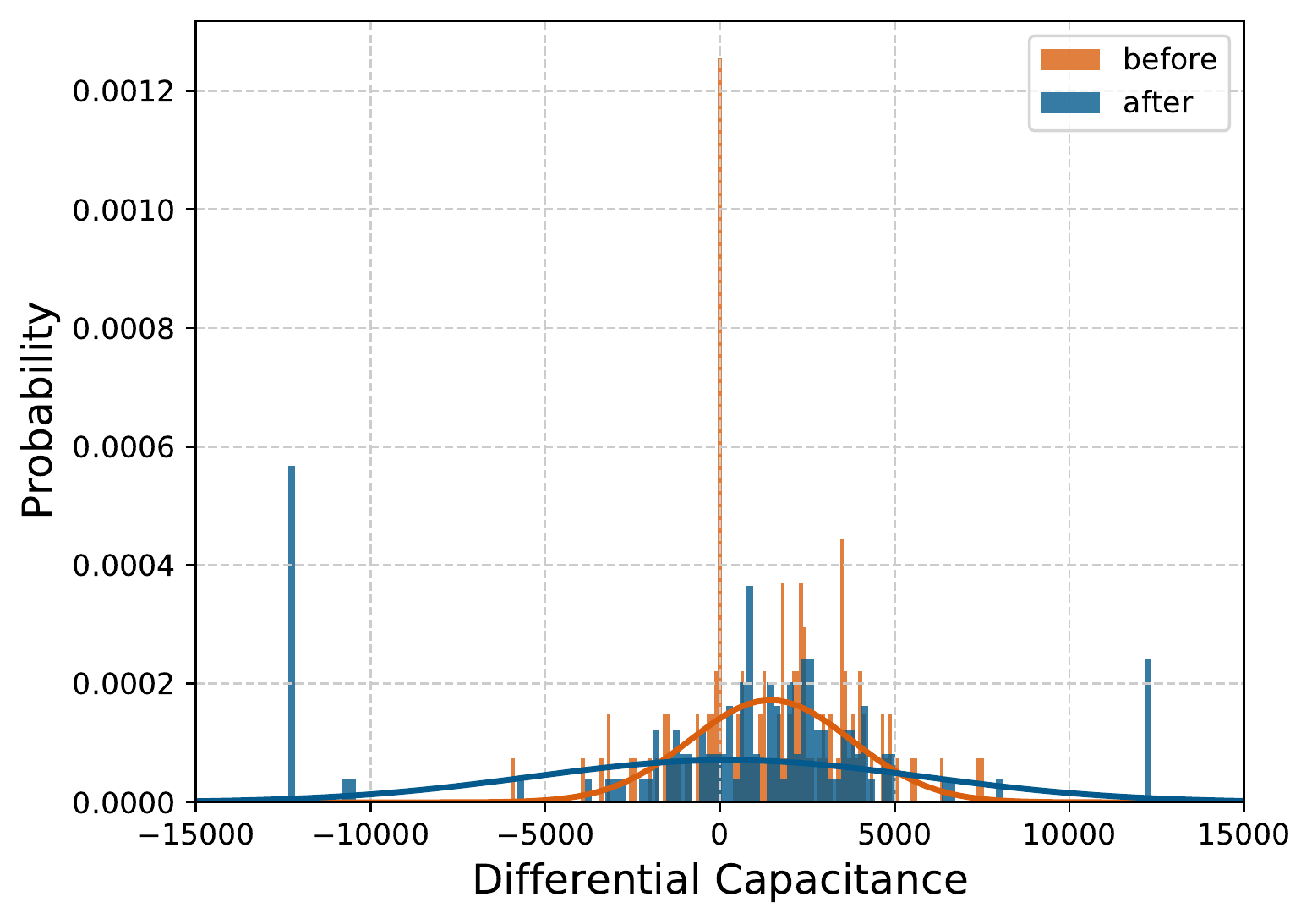}
    \caption{Raw distribution of the differential capacitance (with Gaussian fit) before and after the attack.} 
    \label{fig:attack_hist_raw}
\end{subfigure}\hspace*{0.3cm}%
\begin{subfigure}{.5\textwidth}
    \centering
    \includegraphics[width=\textwidth]{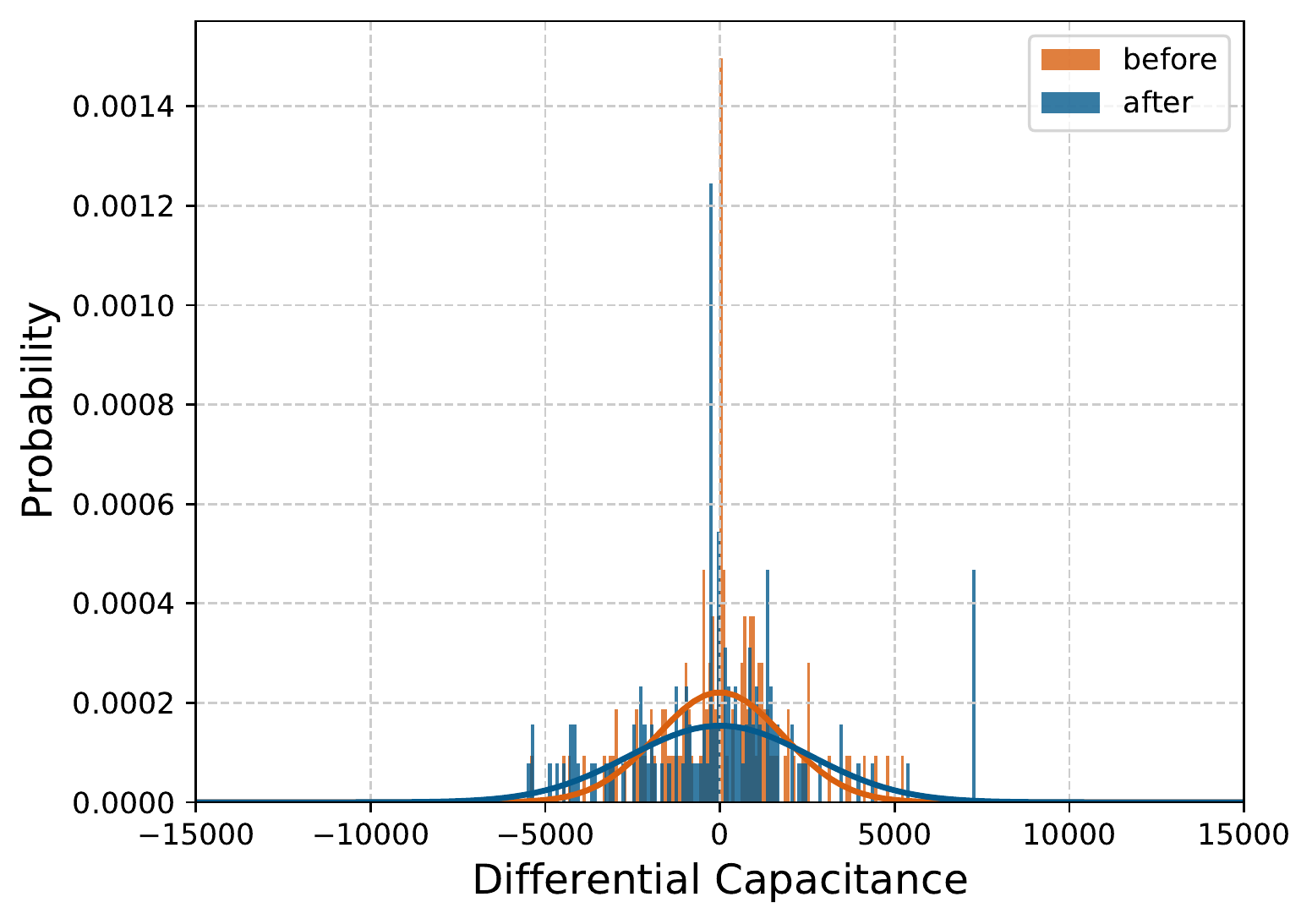}
    \caption{Normalized distribution of the differential capacitance (with Gaussian fit) before and after the attack.}
    \label{fig:attack_hist_normalized}
\end{subfigure}
\caption[]{Analysis of the effect of a drilling attack on the differential capacitance.} 
\end{figure}

The envelope and COVER are designed to withstand drilling attacks with a diameter of \SI{300}{\micro\meter} since both the width and distance of the electrodes are \SI{100}{\micro\meter}.
Hence, a drilling attack destroys at least two electrodes.

As we will see, drilling attacks affect the enclosure in two different ways.
Fig.~\ref{fig:attack_change_first} shows the change in the PUF-response due to such a drilling attack before and after normalization.
The normalization reduces the large offsets and shifts points outside of the $\left[-10000, +10000\right]$ back to the the distribution center.
Values outside of the $\pm 10000$ full-scale range (highlighted in grey) are subjected to non-linear effects (clipping) during the measurement. 
Since, in general, values outside of that range are also possible during a regular measurement, a simple check for values outside of the full-scale range is not sufficient to determine an attack. Furthermore, attacks with smaller drilling diameters might lead to more minor changes in the differential capacitance.
The changes in Figure~\ref{fig:attack_change_first} show that the attack ``muddles up'' 
the differential capacitances of the affected Tx groups. These burst errors have to be considered in the error correction.

Apart from burst errors, drilling attacks also lead to a broadening of the PUF-distribution.
Fig.~\ref{fig:attack_hist_raw} and \ref{fig:attack_hist_normalized} depict the histograms of the PUF-response before and after the attack with and without normalization.
Before the normalization (Fig.~\ref{fig:attack_hist_raw}), the distribution broadens significantly with a change in standard deviation by 3295 points. 
In the case of normalization, the broadening of the distribution is reduced to 787 points since through an attack, a specific Tx group will be more affected than others.
However, it still significantly exceeds the broadening due to temperature changes.

\subsection{Quantization of the PUF-Response}
\label{subsec:quantization}

Quantization is an essential step in post-processing the PUF-response, as discussed in Subsection~\ref{subsec:analysisPUF-response}.
In the following, we discuss previous quantization schemes that were analyzed in the context of the capacitive PUF-based envelope.
Based on the analysis of the PUF in the previous subsections, we determine how different choices of quantization intervals affect the PUF-response.

\subsubsection{Previous Work on Quantization} 
Before analyzing the choice of quantization intervals in more detail, we first give an overview of previous quantization schemes discussed in the context of the PUF-based security enclosure. 
Note that, the quantization does not signify the mapping from continuous to discrete capacitive values in this case.
However, it represents an additional step performed on the already discretized PUF-response --- after the \gls{adc} ---  to reduce the impact of noise during a regular measurement. 

Previous discussions on quantization schemes --- in the the envelope context --- focused on the advantages and disadvantages of equidistant versus equiprobable quantization intervals \cite{IHKS16, IHL+19, LMC, COVER}.
Equiprobable intervals lead to a uniform distribution of the normalized capacitances, while equidistant intervals distribute the capacitances unevenly.
In the case of equiprobable quantization, the analog helper data, stored to shift the PUF-data to the middle of the interval, might leak information about the location of the PUF-data within the distribution.
However, since the helper data reside within the enclosure boundary, they are difficult to access.

Apart from this comparison, previous work also comprised variable-length quantization with equidistant intervals.
This approach maps the PUF-response to binary values of variable length in order to optimize the per-bit minimum entropy \cite{IHL+19}.
The authors tailored the variable-length mapping to \gls{vt} codes, performing a single insertion, deletion, or substitution.

However, variable-length codes are susceptible to increased error propagation in case critical bits or symbols are lost;
this may require synchronization correction and, thus, complicates the decoding process~\cite{GN98, CRR98}.
Another drawback is that a single PUF-node might be distributed over several intervals, which makes its behavior in terms of external influences, like attacks or environmental changes, less predictable.

The choice of an optimal quantization scheme is a non-trivial task, especially for noisy sources \cite{GN98}.
In the following, we aim at providing more clarity on the quantization behavior by investigating the effect of noise and external influences.

\subsubsection{Quantization and Distortion}

The quantization of the PUF-response follows two goals.
On the one hand, the choice of quantization intervals determines the entropy of the PUF-key, and hence, should yield sufficient accuracy.
On the other hand, the intervals should be chosen such that the quantized PUF-response is reliable enough, with a decreased susceptibility to noise.
The optimization of both goals is mutually exclusive and comes with a particular trade-off.
Apart from this trade-off, the quantization also determines the leakage of the PUF-response.
Equidistant intervals leak information about the PUF through varying symbol probabilities, while equiprobable quantization leaks information through the analog helper data. 
Hence, in both cases, we accept an unavoidable information leakage. 

\begin{figure}[b!]
\centering
\hspace{-0.1cm}\begin{subfigure}{.49\textwidth}
    \centering
    \includegraphics[width=\textwidth]{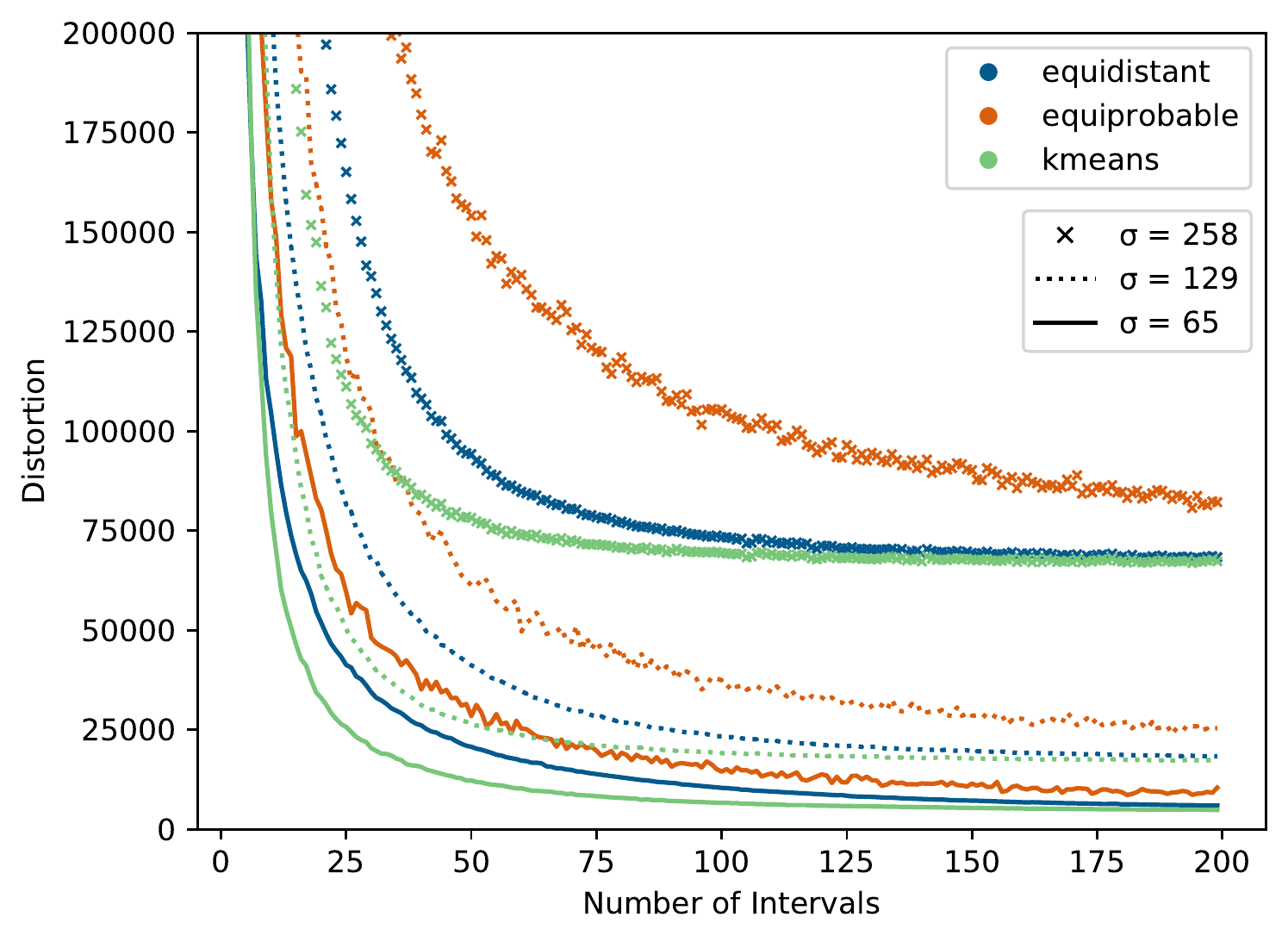}
    \caption{Distortion for equidistant (blue), equiprobable (red), and k-means (green) quantization with three different noise distributions, $\sigma_n=65$, $\sigma_n=129$, and $\sigma_n=258$ for 5 to 200 intervals.}
    \label{fig:distortion}
\end{subfigure}\hspace{0.3cm}%
\begin{subfigure}{.49\textwidth}
    \centering
    \includegraphics[width=0.95\textwidth]{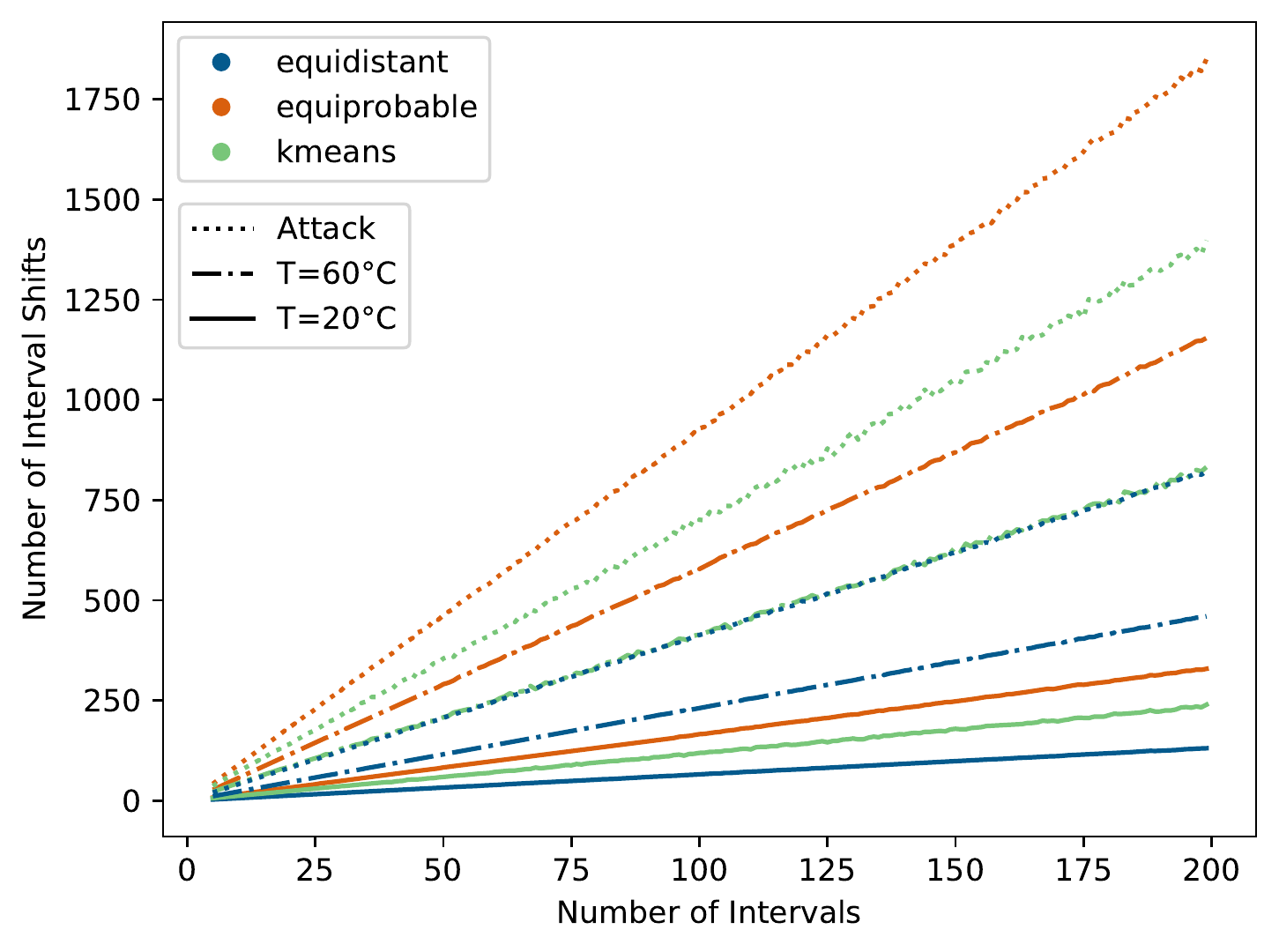}
    \caption{Number of interval shifts in a PUF-response at $T=\ang{20}$, $T=\ang{60}$, and drilling attacks for equidistant (blue), equiprobable (red), and kmeans (green) quantization.}
    \label{fig:number_of_interval_shifts}
\end{subfigure}%
\caption[short]{Distortion (a) and number of changed intervals (b) for equidistant (blue), equiprobable (red), and kmeans (green) quantization.} 
\end{figure}

The quantization quality can be measured as distortion~\cite{GN98} and describes how well the original variable $x$ can be reproduced through the quantization $\hat{x}$.
In the following, we consider a simple distortion measure $d$, the \gls{mse},
$ d(x,\hat{x}) = (x - \hat{x})^2$, which for a sequence of length $n$ reads:
$$ d(x^n,\hat{x}^n) = \frac{1}{n} \sum^n_{i=1} d(x_i,\hat{x_i}) = \frac{1}{n} \sum^n_{i=1} (x_i - \hat{x}_i)^2$$

We simulated the distortion for a set of 1000 noisy PUF-responses, as depicted in Figure~\ref{fig:distortion}, for equidistant (blue), equiprobable (red), and k-means (green) quantization from a total of 5 to 200 intervals.
We see that the distortion will not decrease further at a certain number of intervals. 
This threshold depends on the amount of noise ($\sigma_n = 65$, $\sigma_n = 129$, and $\sigma_n = 258$) that occurs during a regular PUF-measurement.
 
The number of interval shifts for a PUF-response is depicted in Figure~\ref{fig:number_of_interval_shifts}.
We see that, in general, equidistant quantization has a lower probability of interval shifts occuring.
The reason for this is that the innermost intervals of an equiprobable (or k-means) quantization are very narrow, and hence, more minor offsets will have a higher impact compared to equidistant quantization.
We also see that the difference in interval changes between the three cases ($T=\ang{20}$, $T=\ang{60}$, attack) will grow with an increasing number of intervals. 

\begin{figure}[t!]
    \centering
    \includegraphics[width=0.6\textwidth]{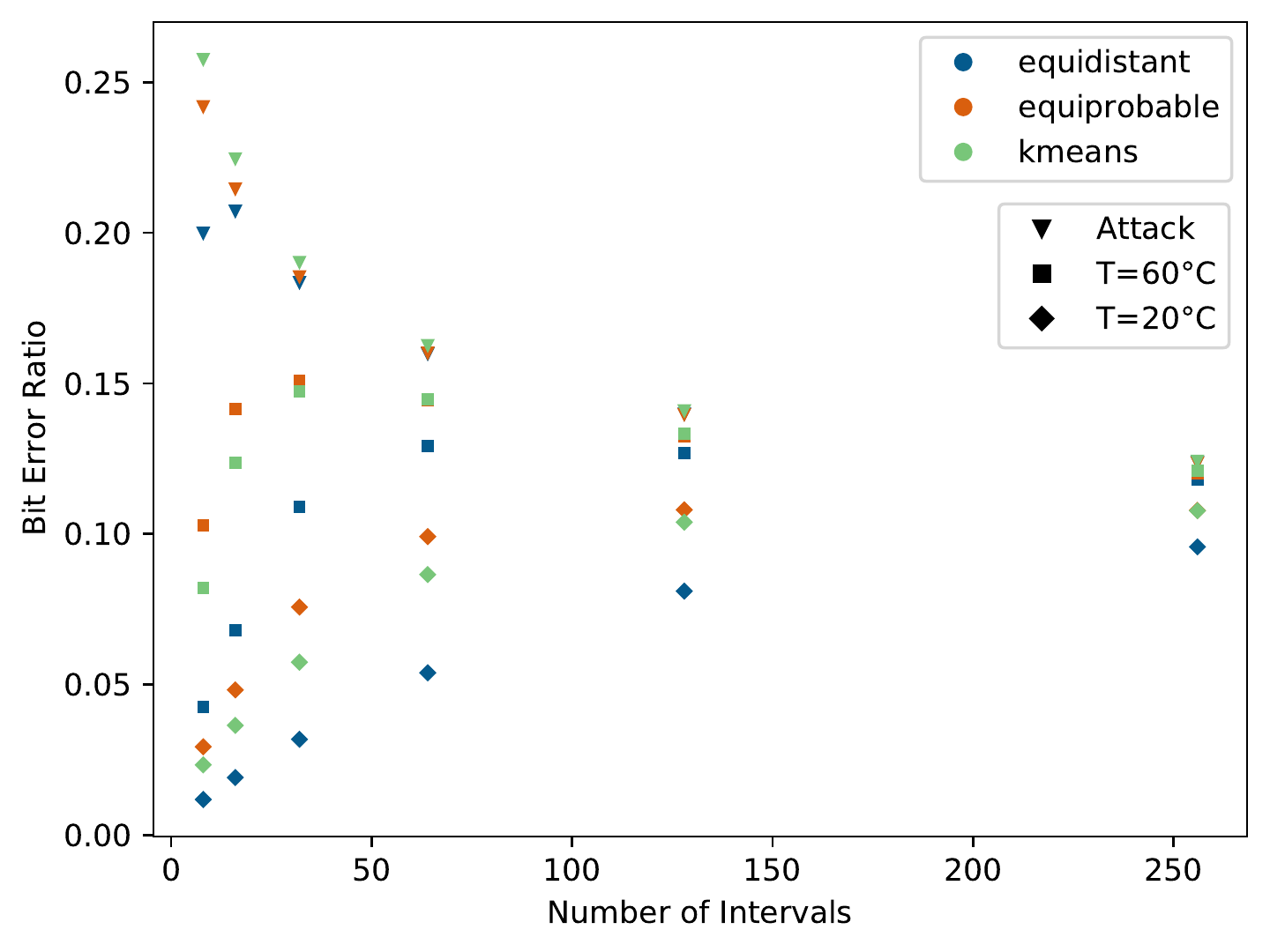}
    \caption{Bit Error Ratio for Gray Code mapping comparing temperatures $T=\ang{20}$ and $T=\ang{60}$ against a drilling attack.} 
    \label{fig:ber}
\end{figure}

Quantized values should be encoded such that the distance between adjacent intervals is minimized. 
For binary numbers, this could be represented by a Gray code mapping.
Figure~\ref{fig:ber} depicts the \gls{ber} for Gray code mapping comparing PUF measurements at $T=\ang{20}$, and $T=\ang{60}$, against a drill attack.
We see that --- for a number of intervals $\le$ 64 --- the \gls{ber} at an elevated temperature can be distinguished from the \gls{ber} of a drill attack.
However, for a large number of quantization intervals, both curves approach the same \gls{ber}. 

This is due to the Gray Code mapping, where each interval is represented by a binary number of $\log_2 (m)$ bits, where $m$ is the number of intervals.
Within the adjacent $\log_2 (m)$ intervals, each interval change --- that corresponds to a bit flip --- increases the Hamming distance of the encoded value by one.
However, when the corresponding PUF-node is changed by more than $\log_2 (m)$ intervals, the Hamming distance decreases again.
Figure~\ref{fig:ber} depicts the bit error rates for Gray encoding with different quantization schemes and temperatures.

Previous assumptions led to a choice of 40 equidistant quantization intervals with a width of 500 points \cite{IHKS16}.
This corresponds to $3.9\sigma_n$, where 99.99\% of values from the noise distribution are within the confidence interval. 
In general, we see that equidistant quantization leads to a lower \gls{ber} and a smaller number of interval changes. 

For 40 intervals, we see that the distortion has not yet reached the threshold, and the difference in interval shifts is still relatively small. 
The distortion for equidistant intervals and k-means reaches the threshold at approximately 100 intervals; however, the \gls{ber} for Gray encoding decreases significantly from 64 intervals to 128. 
A similar behavior can be observed for non-local and balanced Gray Codes and the binary representation of consecutive decimal numbers.
Codes with a larger number of bits, for instance, Johnson code, lead to an even faster decrease of the BER due to a large number of bits per symbol.

Hence, these binary mappings do not describe the error behavior of Figure~\ref{fig:number_of_interval_shifts} accurately enough, and what is more, even benefit the attacker when the number of intervals increases.
In general, modeling the enclosure through a binary channel is difficult.
Thus, we focus on a $q$-ary channel model for our wiretap channel implementation in Section~\ref{sec:wiretapChannel}.

Quantization represents the final step of post-processing. 
The quantized and encoded PUF-response is the input to the key generation, which is discussed in the following.
Note that although we employ a wiretap code to increase security, we still utilize schemes for error correction.

\subsection{Generating Keys from Physical Unclonable Functions}
\label{subsec:keyGen}

In the following, we give an overview of key generation schemes and discuss their suitability for the capacitive PUF-based enclosure. 

\subsubsection{Key Generation Schemes}

When it comes to choosing a key generation scheme, one can either consider pointer-based schemes, where additional reliability information about the \gls{puf} is included, or linear schemes, where a key is generated without considering properties of the \gls{puf}.

In the first case, the advantage of additional reliability comes at the expense of discarding certain PUF-bits.
Applying pointer-based methods to the enclosure leads to ``blind spots''; hence, only reliable parts of the PUF-response constitute the PUF-key.
Thus, the PUF-key becomes less susceptible to changes through an attack, making pointer-based schemes \cite{IBS,Hiller2012,Hiller2013a,YHD15} unsuitable for the capacitive enclosure. 

A variety of linear schemes do not consider reliability information about the PUF and, hence, derive a key based on the complete PUF-response.
One of the first schemes proposed for \glspl{puf}, the Fuzzy Commitment \cite{fuzzyCommitment}, which was extended by adding the quantization and post-processing steps for the capacitive enclosure, is depicted in Figure~\ref{fig:keyGeneration}. 

It consists of two major steps:
(i) During the enrollment at the manufacturer, a key is chosen and the corresponding helper data are generated;
to be more precise, a true random number $R$ generates the secret $S$, which is then encoded to the codeword $C$.
(ii) In the reproduction phase, the key is repeatedly restored through measuring the PUF-response during the regular operation of the protected device.

In the enrollment phase, the measured and normalized PUF-response $X$ is quantized $q(x)$, and the analog helper data $W'$ are generated and stored in NVM.
From the codeword together with the quantized PUF-response $\widehat{X}$, the helper data $W$ are calculated and stored in NVM to end the enrollment.
The ``channel'' is obtained by repeatedly reproducing the PUF-reponse  
that is subjected to noise $\epsilon_n$, environmental effects like temperature changes $\epsilon_t$, and possible drilling attacks $\epsilon_a$. 
This can be seen as a ``faulty'' codeword ``transmitted'' over a noisy channel, 
that when decoded during reconstruction, yields the secret $S$.
In order to verify the secret, which is a \gls{kek}, the key chain is decrypted.
If this is successful, the secret $S$ is valid. Otherwise, the alarm and zeroization are triggered.
\footnote{The Fuzzy Commitment scheme is not limited to binary linear codes only, but e.g. Reed Solomon codes (e.g. with $q-$ary symbols) are also possible \cite{chauhan2016fuzzy}.
This also extends to the $q-$ary polar code in Section \ref{subsec:enc_dec}.} 

The Fuzzy Extractor \cite{fuzzyExtractor} is another scheme, which is similar to the Fuzzy Commitment. In this case, however, the PUF-response and not a true random number constitutes the secret, which is why the secret is hashed in order to reduce helper data leakage.

A scheme, where no random number from a \gls{trng} is generated, is the Syndrome Construction \cite{fuzzyExtractor, Dodis2008}. Just as in case of the Fuzzy Extractor, the secret is equivalent to the PUF-response, which makes additional hashing necessary. 

In the basic Syndrome Construction, the helper data $W = \widehat{X}\,H^\mathrm{T}$ are defined via the parity check matrix $H$. The reconstruction consists of minimizing the error $e$ in $W = (\widehat{X}+e)\,H^\mathrm{T}$.
In \cite{CIW+17}. This approach is implemented in the context of polar codes, however, as we will see in Section~\ref{sec:wiretapChannel} it is not suitable for wiretap coding. 

Another scheme that does not require an additional random number $R$ is \gls{sllc} \cite{Hiller2015a}. In this case, the PUF-response $X = X_\mathrm{S} + X_\mathrm{M}$ is split into the secret $X_\mathrm{S}$ and a part $X_\mathrm{M}$ for masking. The helper data is defined as $W=X_\mathrm{S}P\oplus X_\mathrm{M}$. \gls{sllc} has the drawback that hashing is required. Furthermore, it can only be applied to systematic codes, where the information and redundancy are separated.

A key generation scheme based on the secure sketch that does not require additional helper data was proposed by Müelich and Bossert \cite{MB17}. The \gls{hda} is necessary to create a codeword from the PUF-response, however, Müelich and Bossert constructed the code, such that the PUF-response corresponds to a codeword. A drawback that the authors described was the increased complexity of their scheme.

\begin{figure}[t!]
\centering
\footnotesize
\input{secs/fuzzy_commitment}
 \normalsize
 \caption{Key generation via Fuzzy Commitment including post-processing and quantization.}
\label{fig:keyGeneration}
\end{figure}

For the capacitive PUF-based enclosure, we choose the Fuzzy Commitment as depicted in Figure~\ref{fig:keyGeneration} due to two main reasons:
(i) This allows for a full flexibility regarding the key. Because the key is generated from a TRNG, a second enrollment is possible e.g. after transport (see \cite{FORTRESS}).
Besides, no hash function is required.
(ii) As discussed in Section~\ref{ssec:code_construction}, it is necessary for the wiretap setting to introduce randomness. This is not possible for key generation schemes such as Syndrome constructions or \gls{sllc} because the key and helper data only depend on the PUF response without additional randomness.

As the overall key generation scheme has been chosen, the next step is to find a suitable encoder and decoder for the error correction.

\subsubsection{Previous Error Correction Codes}

Several error correction codes were proposed for the enclosure, for instance, Reed Solomon, \gls{bch}, or \gls{vt} codes \cite{LMC, IHL+19}.
However, a major issue regarding error correction in the context of the enclosure is that the chosen code might correct changes in the PUF-response that originate from an attack. Many of the previously proposed schemes do not address this problem.

To tackle this issue, error correction based on Limited Magnitude Codes (LMCs) was proposed \cite{LMC, originalLMC}, where only the \glspl{lsb} of each of the PUF-symbols are corrected. Through this, the error correction is restricted to the neighboring intervals, leaving larger offsets caused by an attack unchanged. 
In general, \glspl{lmc} provide a simple method to incorporate tamper-sensitivity without considering complex error patterns.

In order to describe the more complex patterns of burst errors and changes in the PUF-distribution, we propose a wiretap channel implementation based on polar codes. By choosing Polar Codes, we are able to model both effects --- environmental influences and attacks --- separately, as we will see in the next section.
This separation requires a wiretap code and is not feasible with regular PUF error correction codes. 

%% file: secs/post-processing.tex
\tikzstyle{block} = [rectangle, draw, text width=15em, text centered, minimum 		width=15em, minimum height=2em]
\tikzstyle{blockDiagonal} = [rectangle with diagonal fill, draw, text width=7em, text centered, minimum width=7em, minimum height=3em]
\tikzstyle{line} = [draw, -, thick]
\tikzstyle{arrow} = [draw, -latex, thick]
\tikzstyle{block} = [rectangle, draw, text centered, minimum height=2em]
\definecolor{lightergrey}{rgb}{0.83, 0.83, 0.83}
\pgfmathdeclarefunction{gauss}{2}{%
  \pgfmathparse{1/(#2*sqrt(2*pi))*exp(-((x-#1)^2)/(2*#2^2))}%
}
\pgfplotsset{%
    width=.45\textwidth,
}
\begin{figure}[b!]
\centering
\footnotesize
\begin{scaletikzpicturetowidth}{\textwidth}
\begin{tikzpicture}[scale=0.6]
\begin{axis}[
    axis x line=center,
  	axis y line=center,
  	xtick={0},
  	ytick={0},
    xmin = -12000, xmax = 12000,
    ymin = 0.0, ymax = 0.0002
    ]
    \addplot[
    	mark=none,
        domain = -10000:10000,
        samples = 200,
        smooth,
        thick,
        blue,
        width = 0.2\textwidth,
        height = 0.1\textwidth,
        xlabel = {$x$},
    	ylabel = {$y$},
    ] {gauss(1000, (2241+600))}; 
    \node at (axis cs:0,0.00025) {\Large measurement};
\end{axis}
\path [arrow] (4.7,2) -- (5.5,2);
\node at (2.3,-0.5) {measurement};

\end{tikzpicture}%
\begin{tikzpicture}[scale=0.6]
\begin{axis}[
    axis x line=center,
  	axis y line=center,
  	xtick={0},
  	ytick={0},
    xmin = -10000, xmax = 10000,
    ymin = 0.0, ymax = 0.0002
    ]
    \addplot[
    	mark=none,
        domain = -10000:10000,
        samples = 200,
        smooth,
        thick,
        dashed,
        blue,
        width = 0.2\textwidth,
        height = 0.1\textwidth,
        xlabel = {$x$},
    	ylabel = {$y$},
    ] {gauss(1000, (2241+600))}; 
   \addplot[
    	mark=none,
        domain = -10000:10000,
        samples = 200,
        smooth,
        thick,
        blue, 
        width = 0.2\textwidth,
        height = 0.1\textwidth,
        xlabel = {$x$},
    	ylabel = {$y$},
    ] {gauss(0, (2241))};
    \node at (axis cs:0,0.00025) {\Large normalization};
\end{axis}
\path [arrow] (4.7,2) -- (5.5,2);
\node at (2.3,-0.5) {normalization};

\end{tikzpicture}
\begin{tikzpicture}[scale=0.6]
\begin{axis}[
    axis x line=center,
  	axis y line=center,
  	xtick={0},
  	ytick={0},
    xmin = -10000, xmax = 10000,
    ymin = 0.0, ymax = 0.0002
    ]
    \addplot[
    	mark=none,
        domain = -10000:10000,
        samples = 200,
        smooth,
        thick,
        blue,
        width = 0.2\textwidth,
        height = 0.1\textwidth,
        xlabel = {$x$},
    	ylabel = {$y$},
    ] {gauss(0, (2241))}; 
    \node at (axis cs:0,0.00025) {\Large quantization};
\end{axis}
\path [line, dashed] (0,0.0) -- (0,4.5);
\path [line, dashed] (1.2,0.0) -- (1.2,4.5);
\path [line, dashed] (1.9,0.0) -- (1.9,4.5);
\path [line, dashed] (2.7,0.0) -- (2.7,4.5);
\path [line, dashed] (3.4,0.0) -- (3.4,4.5);
\path [line, dashed] (4.4,0.0) -- (4.4,4.5);
\path [arrow] (4.7,2) -- (5.5,2);
\node at (2.3,-0.5) {quantization};
\end{tikzpicture}
\begin{tikzpicture}[scale=0.6]
\begin{axis}[
    axis x line=center,
  	axis y line=center,
  	xtick={0},
  	ytick={0},
    xmin = -10000, xmax = 10000,
    ymin = 0.0, ymax = 0.0002
    ]
    \addplot[
    	mark=none,
        domain = -10000:10000,
        samples = 200,
        smooth,
        thick,
        blue,
        opacity=0.4,
        width = 0.2\textwidth,
        height = 0.1\textwidth,
        xlabel = {$x$},
    	ylabel = {$y$},
    ] {gauss(0, (2241))};
    \node at (axis cs:0,0.00025) {\Large quantization};
\end{axis}
\node at (0.6, 0.01)[circle,fill,color=blue,inner sep=1.5pt]{};
\node at (1.55, 1.25)[circle,fill,color=blue,inner sep=1.5pt]{};
\node at (2.275, 3.25)[circle,fill,color=blue,inner sep=1.5pt]{};
\node at (3, 1.25)[circle,fill,color=blue,inner sep=1.5pt]{};
\node at (3.95, 0.01)[circle,fill,color=blue,inner sep=1.5pt]{};
\path [line, dashed] (0,0.0) -- (0,4.5);
\path [line, dashed] (1.2,0.0) -- (1.2,4.5);
\path [line, dashed] (1.9,0.0) -- (1.9,4.5);
\path [line, dashed] (2.65,0.0) -- (2.65,4.5);
\path [line, dashed] (3.35,0.0) -- (3.35,4.5);
\path [line, dashed] (4.55,0.0) -- (4.55,4.5);
\node at (2.3,-0.5) {analog helper data};
\end{tikzpicture}
\end{scaletikzpicturetowidth}
\normalsize
\caption{Post-processing of the PUF-response after measurement and helper data generation.}
\label{fig:post-processing}
\end{figure}

%% file: secs/fuzzy_commitment.tex
\tikzstyle{block} = [rectangle, draw, text width=15em, text centered, minimum 		width=15em, minimum height=2em]
\tikzstyle{blockDiagonal} = [rectangle with diagonal fill, draw, text width=7em, text centered, minimum width=7em, minimum height=3em]
\tikzstyle{line} = [draw, -, thick]
\tikzstyle{arrow} = [draw, -latex, thick]
\tikzstyle{block} = [rectangle, draw, text centered, minimum height=2em]
\definecolor{lightergrey}{rgb}{0.83, 0.83, 0.83}

\begin{tikzpicture}
\draw[draw=black, fill=lightergrey, opacity=0.6] (-4.3,-3.85) rectangle (0.6,0.4);
\node [draw=none, fill=none, xshift=-2.9cm, yshift=-0.05cm](keyEnrollment){\textcolor{darkgray}{\textbf{Key enrollment}}};
\draw[draw=black, fill=lightergrey, opacity=0.6] (3.3,-3.85) rectangle (9.05,0.4);
\node [draw=none, fill=none, xshift=7.4cm, yshift=-0.05cm](keyReproduction){\textcolor{darkgray}{\textbf{Key reproduction}}};
\draw[dashed] (-0.7,-5.1)--(-0.7,-0.7);
\draw[dashed] (4.65,-5.1)--(4.65,-0.7);
\node [draw=none, fill=none](originPUF){$X$};
\node [block, below of=originPUF, node distance=1.3cm](quantization){$q(X)$};
\node [draw=none, fill=none, right of=quantization, node distance=2.0cm](analogHelperData){$W'$};
\node [block, below of=analogHelperData, font=\scriptsize, fill=lightergrey, draw=lightergrey, node distance=0.6cm] (nvm) {NVM};

\node [draw=none, fill=none, below of=quantization](quantizedPUF){$\widehat{X}$};
\node [draw=none, fill=none, left of=quantizedPUF, below of=quantizedPUF, xshift=-0.3cm](codeword){$C$};
\node [block, left of=codeword](enc){enc};
\node [draw=none, fill=none, left of=enc, node distance = 1.4cm](trng){$R\equiv S$};
\node [draw=none, fill=none, below of=quantizedPUF](xor){$\oplus$};
\node [draw=none, fill=none, right of=xor, node distance=2.0cm](helperData){$W=C\oplus \widehat{X}$};
\node [block, below of=helperData, font=\scriptsize, fill=lightergrey, draw=lightergrey, node distance=0.6cm] (nvm1) {NVM};

\node [draw=none, fill=none, right of=helperData, right of=helperData](xor1){$\oplus$};
\node [draw=none, fill=none, right of=helperData, right of=helperData, node distance=2.0cm](codeword2){\hspace{0.25cm}\makecell[l]{$C+\hat{\epsilon}_{n}+$\\$+\hat{\epsilon}_{t}+\hat{\epsilon}_{a}$}};
\node [block, right of=codeword2, node distance=1.7cm](dec){dec};
\node [draw=none, fill=none, above of=xor1](quantizedPUF2){$\widehat{X}'$};
\node [block, above of=quantizedPUF2](quantization2){$q(\widetilde{X})$};
\node [draw=none, fill=none, above of=quantization2, node distance=1.1cm](reproducedPUF){\hspace{0.25cm}\makecell[l]{$X+\epsilon_{n}+$\\$+\,\epsilon_{t}+\epsilon_{a}$}};
\node [draw=none, fill=none, right of=dec](secret){$S$};
\node [draw=none, fill=none, below of=helperData, node distance=1.7cm](bsc){noisy channel};
\node [draw=none, fill=none] at ([yshift=0cm, xshift=-0.5cm] codeword2) (aux){};

\path [arrow] (trng) -- (enc);
\path [arrow] (enc) -- (codeword);
\path [arrow] (codeword) -- (xor);
\path [arrow] (originPUF) -- (quantization);
\path [arrow] (quantization) -- (analogHelperData);
\path [arrow] (quantization) -- (quantizedPUF);
\path [arrow] (quantizedPUF) -- (xor);
\path [arrow] (xor) -- (helperData);
\path [arrow] (helperData) -- (xor1);
\path [arrow] (xor1) -- (aux);
\path [arrow] (codeword2) -- (dec);
\path [arrow] (dec) -- (secret);
\path [arrow] (reproducedPUF) -- (quantization2);
\path [arrow] (quantization2) -- (quantizedPUF2);
\path [arrow] (quantizedPUF2) -- (xor1);
\path [arrow] (analogHelperData) -- (quantization2);
\end{tikzpicture}

%% file: secs/wiretap.tex
\section{Implementing the Wiretap Channel}
\label{sec:wiretapChannel}

The applied error correction code has to be able to distinguish between environmental changes and an attack.
A theoretical model that incorporates this separation between two different channels, is the wiretap channel \cite{WynerI, WynerII, Csiszar}. 
In the following, we model the error behavior of the enclosure with the help of the wiretap channel, and propose an implementation based on polar codes. 

\subsection{The Wiretap Channel and Attacker Model}

The wiretap channel originally regards the problem of a wiretapper eavesdropping on a discrete, memoryless channel \cite{WynerI}. Wyner showed that a reliable transmission with a finite capacity is possible, while achieving approximately perfect secrecy. For this, he assumed that the wiretapper eavesdrops the transmission via a second channel. Hence, the goal is to provide a reliable transmission on the main channel, hiding it from the second channel accessed by the wiretapper.
In general, there exist different assumptions for error probability and secrecy \cite{WynerII}.

\begin{figure}[b!]
\centering
\footnotesize
\input{secs/wiretap_channel_pic}
 \normalsize
\caption{A wiretap channel adaptation for measurement noise $\hat{\epsilon}_n$, environmental changes $\hat{\epsilon}_t$ and drilling attacks $\hat{\epsilon}_a$.} 
\label{fig:wiretap}
\end{figure}
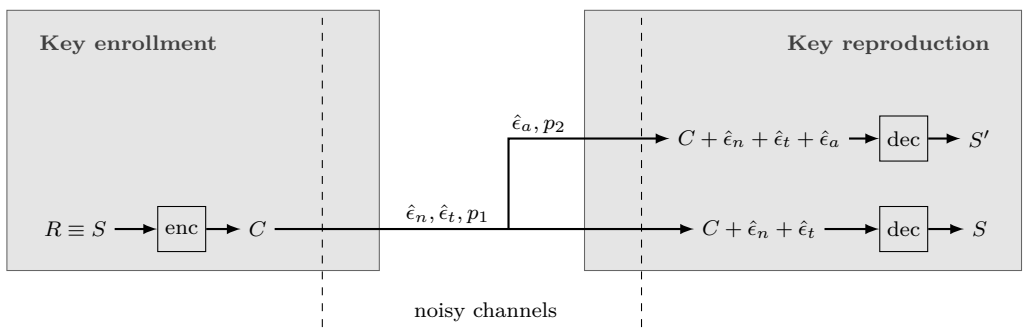

The separation into a legitimate and a noisy channel can be applied to the capacitive PUF-based enclosure in order to improve the error correction. 
We adapt the wiretap channel for our purposes, see Figure~\ref{fig:wiretap}, such that on the main channel, the codeword $C = \widehat{X} \oplus W$ is ``transmitted'' with error probability $p_1$, considering changes in the PUF-distribution that stem from noise $\hat{\epsilon}_n$ and temperature changes $\hat{\epsilon}_t$. On the second channel, the codeword is additionally affected by changes due to an attack $\hat{\epsilon}_a$, which leads to a different error probability $p_2$.

Previous polar code implementations of wiretap channels in the PUF-context focused on binary silicon \glspl{puf} and hiding secrecy leakage from unstable or biased PUF-bits \cite{hiller2017hiding, BY19, Bai2021}. These wiretap codes dissected their implementation into a regular channel and a channel where the distorted helper data is transmitted. Our use case and code construction significantly differ from previous implementations since the helper data do not have any relevance in our case. We construct the wiretap channel from a channel modeling the regular transmission of the PUF-response, while the other channel models the PUF-response under attack, e.g., through a drilling attack. What is more, our code construction is based on $q$-ary polar codes.
To the best of our knowledge, we are, hence, the first to provide a wiretap channel design and implementation with higher-order alphabet PUFs targeting physical layer security.

Regarding $\hat{\epsilon}_a$, we define our attacker to have two possibilities:
(i) When the overall device enclosed by the PUF is powered off, the attacker can remove the envelope. He can then try to decrypt the data that has been secured by the PUF key.
(ii) During runtime of the device, he can drill a hole into the envelope and try to gain access to the sensitive data in a minimally invasive way to prevent zeroization.

An investigation of micro-drilling attacks, magnetic probing, and bypassing of electrodes was published by Garb et al.\ \cite{GSHO21} in the context of the capacitive PUF-based enclosure, including countermeasures. The enclosure was designed \cite{COVER, B-TREPID} to withstand 250-\SI{300}{\micro\meter} diameter drilling attacks that can be reliably detected. Smaller holes are achieved through laser beams with high aspect ratios. \glspl{fib} are, in general, more suited for processing surface structures. An attempt to drill a deep hole with \glspl{fib} will lead to debris at the bottom of the hole due to the removed material, making small holes with a diameter of approximately \SI{10}{\micro\meter} infeasible. However, with lasers and high aspect ratios, small holes in the two-digit micrometer range are, in general, achievable. The thickness of the casing and an additional potting material hinder an attacker from reaching the components on the \gls{pcb} since a deeper hole comes with a larger diameter. Another factor is the probing of critical components, which is limited by the shaft width of the probing needle.
The tip length of the smallest commercially available probing needles amounts to 3.3-\SI{5}{\milli\meter}, while the shaft width is usually in the higher two-digit micrometer range and will, hence, not fit through an arbitrarily small hole. The reach of the probing needle can be further limited by increasing the casing thickness.
Since small holes, in general, require additional countermeasures that can not be achieved through error correction, we focus on detecting holes with a diameter of 250-\SI{300}{\micro\meter}. 
The envelope provides good protection against fault injection since it covers the entire device. The only component that can be externally accessed is the voltage supply. However, voltage glitching is mainly hampered by monitoring the voltage supply. Also, hardware-based countermeasures through an optimized choice of electronic components make voltage glitching attacks practically infeasible.
When switching out the PUF-based solution against a battery-backed system, the power supply will reside within the boundary of the enclosure, leaving no attack point for \gls{lfi} or glitching attacks. However, even in this case, faults induced by radiation are still conceivable and have to be counteracted by extensive software hardening.

In the following, we provide an implementation of the wiretap channel through $q$-ary polar codes such that in either attack scenario, (i) and (ii), security is achieved.
Our code construction can be applied to other single-challenge PUFs in the context of error correction. In our system model, the PUF serves as a \gls{kek} that protects the other critical security parameters in the key chain. Also, device authentication can be achieved by creating additional key pairs~\cite{FORTRESS}.

\subsection{Polar Codes}

Polar codes were introduced by Stolte \cite{StoltePhD} and later reintroduced by Arıkan \cite{Arikan}. Arıkan showed that polar codes achieve the capacity of a binary-input \gls{dmc} under \gls{scd} decoding. Later works extended Arıkan's result to polar codes defined over a binary extension field, i.e., polar codes with symbols from the extension field $\mathbb{F}_q$, where $q$ is a power of two. Yuan and Steiner introduced the construction of polar codes using a kernel defined over an extension field \cite{Yuan_Steiner}.
Besides, polar codes also provide a capacity-achieving construction for the wiretap channel with symmetric component channels~\cite{polar_wiretap}.

Unlike previous polar codes for \glspl{puf} \cite{hiller2017hiding, BY19, Bai2021}, we focus on a generalized $q$-ary polar code to extract more entropy and a longer secret from the \gls{puf}. Our analysis in Subsection~\ref{subsec:quantization} showed that modeling the PUF-response with a binary symmetric channel led to small changes in the PUF-response from a tampering attempt. Thus, to use a valid wiretap coding scenario that guarantees a higher security level, we need to shift to a higher quantization level. The fact that the PUF-response is analog motivates even further the usage of a quantization process with more than just two levels. The advantages of a $q$-ary model for the enclosure are an increased sensitivity towards tampering and a higher extraction of entropy.
In the following, we discuss the encoder and decoder for the capacitive PUF-based enclosure and the code construction.

\subsubsection{Encoder and Decoder}
\label{subsec:enc_dec}

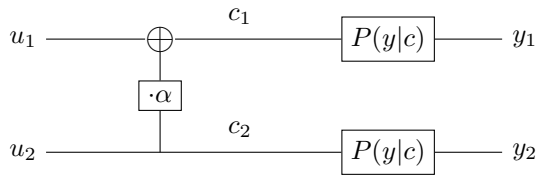
\begin{figure}[b!]
    \centering
    \input{kernel}
    \caption{Polar code $2 \times 2$ kernel defined over a $q-$ary field. Note that $\alpha$ is an element of the field $\mathbb{F}_q$.}
    \label{fig:q_ary_kernel}
\end{figure}

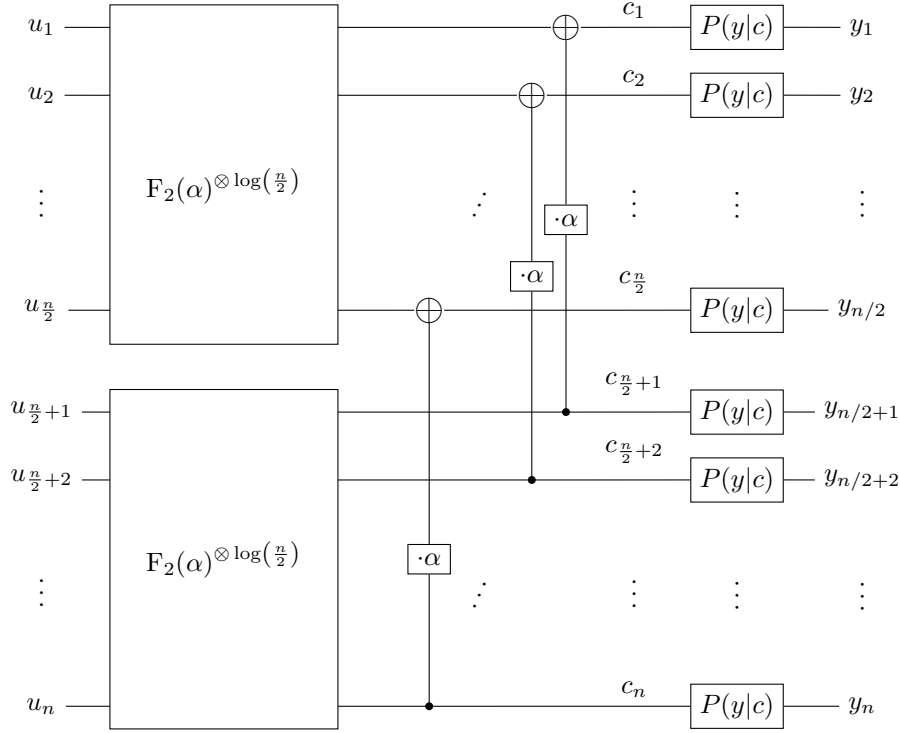
\begin{figure}[t!]
    \centering
    \input{polar_code_figure}
    \caption{Polar encoder}
    \label{fig:encoding}
\end{figure}

Figure~\ref{fig:q_ary_kernel} defines a polar code over a finite field $\mathbb{F}_q$~\cite{Yuan_Steiner}. Similarly to binary polar codes~\cite{Arikan}, $q$-ary polar codes are based on a process called \emph{channel polarization}. Let the source output be two $q-$ary symbols $(u_1, u_2)$ and let their encoded version be two $q-$ary symbols $(c_1, c_2)$. Then the relation between those two vectors is 
\begin{equation}\label{eq:first_pol}
(c_1, c_2) = (u_1, u_2) \cdot \begin{pmatrix} 1 & 0 \\
\alpha & 1 
\end{pmatrix} \triangleq (u_1, u_2)\cdot \text{F}_2(\alpha)
\end{equation}
where $\alpha$ is an optimization parameter~\cite{Yuan_Steiner}, $\text{F}_2(\alpha)$ is the polarization matrix, which in the $q-$ary case depends on $\alpha$. As illustrated in Figure~\ref{fig:q_ary_kernel}, two copies of the same physical channel (defined by the channel law $P(y\lvert c)$) are polarized into two virtual channels. The first channel with an input $u_1$ and an output $(y_1, y_2)$ has a lower rate than the physical channel, and the second channel that inputs $u_2$ and outputs $(y_1,y_2,u_1)$ has a higher rate than the physical channel~\cite{Arikan}. This case can be generalized for a code length $n > 2$, as depicted in Figure~\ref{fig:encoding}. 
The information symbols $u = (u_1, u_2, u_3, u_4, \dots, u_n)$ are encoded using a polar transform that yields the codeword $c = (c_1, c_2, c_3, c_4, \dots, c_n)$. 
The relation between $u$ and $c$ can be written as
\begin{equation}
    c = u\cdot \mathrm{F}_2(\alpha)^{\otimes \log_2 n}
\end{equation}
where $\mathrm{F}_2(\alpha)^{\otimes \log_2 n}$ denotes the $\log_2 n-$fold Kronecker product of the matrix $\mathrm{F}_2(\alpha)$ with itself.
The channel output is denoted by $y = (y_1, y_2, y_3, y_4, \dots, y_n)$, which represents the erroneous observation of $c$.
Similarly, $n$ physical channels will be polarized into channels with a rate higher than the physical one, and into channels with a lower rate than the physical channel ~\cite{Arikan}.
Arıkan showed that with $n \rightarrow \infty$, the channel polarization results in either perfect or useless channels, i.e., channels with capacity $1$ or $0$. These results were also later generalized to the $q-$ary case~\cite{polar_codes_q_ary_capacity}. However, for short or moderate code length, the channel polarization is far from the asymptotic one, thus, allowing some channels to be considered as \emph{mediocre}, i.e., neither perfect nor useless. The aim of the code construction is to find the best polarized channels and transmit information over those and freeze the other channels, i.e., transmit a predetermined \emph{frozen} value over the bad channel that contains no information. We will use the symbol $0$ as a frozen value and we denote the frozen positions by the set $\mathcal{F}$.

\begin{figure}[t!]
    \centering
    \input{polar_decoder}
    \caption{Polar decoder}
    \label{fig:decoding}
\end{figure}
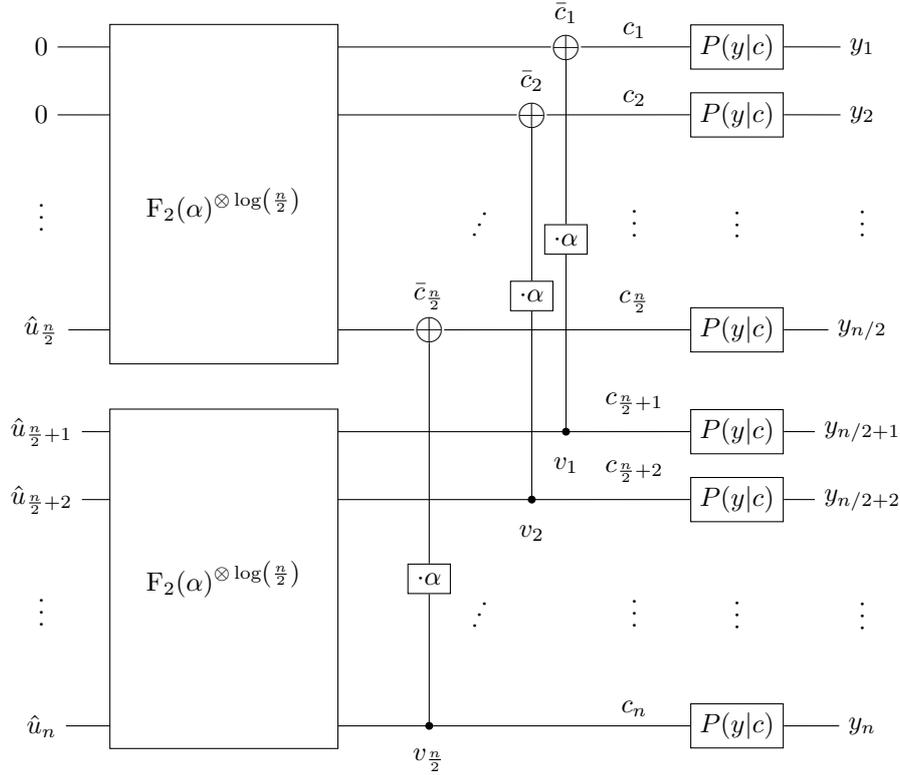

The decoder is depicted in Figure~\ref{fig:decoding}. 
The task of the decoder is to determine an estimate $\hat{u} = (\hat{u}_1, \hat{u}_2, \hat{u}_3, \hat{u}_4, \dots, \hat{u}_n)$ of the source vector, knowing the received $y$ and the frozen positions $\mathcal{F}$ and their values. \gls{scd} decoding is the original decoding algorithm that Arikan proposed for polar codes under which they achieve capacity~\cite{Arikan}.

\gls{scd} decoding is a decoding strategy that allows the use of soft information and can be implemented recursively due to the structure of polar codes. The main idea is that the soft information is propagated from the right to the left (see Figure~\ref{fig:decoding}) and \emph{check node} operations are performed (depicted with an XOR operation and iterated as $\tilde{c}_i$ in Figure~\ref{fig:decoding}). Then after the soft information has reached the very top left positions (the positions of $\hat{u} = (\hat{u}_1, \hat{u}_2, \dots, \hat{u}_n)$ in Figure~\ref{fig:decoding}), a hard decision must be performed by taking into consideration the soft information available. Then, the hard decision is propagated from left to right
, thus, allowing the decoder to successively decide on the bits that were most likely transmitted. The hard decision is then ``mixed'' with the soft information for the \emph{variable node} operation (depicted with a bullet point and called $v_i$ in Figure~\ref{fig:decoding}). 

Moreover, since the decoder knows the positions and the values of the frozen bits, it can bypass the hard estimation for those particular positions and ``decide'' on the correct information. However, hard decisions can mislead to a miscorrected symbol, thus allowing errors to be propagated for later hard decisions. This issue was solved by Tal and Vardy in their work in~\cite{Tal2011}, where they introduced \gls{scl} decoding. 

\gls{scl} decoding allows the decoder to pursue many choices for an information position once it needs to make a hard decision for a non-frozen symbol. However, to allow a practically feasible decoder, a certain list size $L$ was introduced. Once the list of paths has been filled up (when $L$ possible paths are stored in the list stack), the decoder needs to discard the most unlikely paths and only pursue the most likely ones. The work in~\cite{Tal2011} showed that even the smallest possible $L=2$ decoder improved the performance compared to \gls{scd} decoding and almost maximum likelihood decoding performance was achieved by a relatively small list size $L = 32$ for some parameters~\cite{Tal2011}. For the $q-$ary case, the authors in~\cite{Yuan_Steiner} proposed a pruned \gls{scl} decoder that only pursues the paths that are within a threshold $\delta$ of reliability. The main difference to conventional \gls{scl} decoding is that the list does not have to be full to start pruning. The performance is very comparable to the conventional \gls{scl} decoding, but the number of operations is significantly reduced in the case where transmission occurs over a good channel~\cite{Yuan_Steiner}.

Tal and Vardy \cite{Tal2011} showed that indeed polar codes suffer from a bad weight distribution compared to other state-of-the-art codes since many times the transmitted codeword was part of the final decoding list but was not the most likely to be picked. Thus, a modification of the decoding is needed such that it picks --- with a high probability --- the transmitted codeword from the final list (if it is a member of it). This can be achieved by a code concatenation with CRC~\cite{Tal2011} or by deploying a hash computation of all the candidates of the final list~\cite{CIW+17}.

\subsubsection{Code Construction}
\label{ssec:code_construction}

The aim of our code construction is to determine the polarized channels that are ``bad'' for the attacker channel and ``good'' for the legitimate channel affected by noise and temperature changes. We estimate the quality of the polarized channels by using the Monte-Carlo code construction~\cite{Arikan} for both types of channels (legitimate and eavesdropper). However, since we will operate on a short length regime (the length of the polar code $n$ is much smaller than infinity), we cannot have fully polarized symbol channels.
In~\cite{polar_wiretap}, a polar code construction is presented that achieves the secrecy capacity of a wiretap channel when the channel components are symmetric. This polar code construction allows the polarization of the symbol-channels $u_i \leftarrow y_1, y_2, \dots, y_n, u_1,\dots u_{i-1}$ for the channels of the legitimate receiver and the eavesdropper. Then, the polarized symbol-channels that are ``very bad'' are frozen, i.e., the input symbol is set to a fixed value, the polarized symbol-channels that are ``good'' for the legitimate receiver but ``bad'' for the eavesdropper are used to transmit information symbols and finally, the polarized symbol channels that are ``good'' for both receivers are used to transmit random symbols that carry no information.

In order to estimate the quality of the polarized symbol-channels we need a realistic channel model for both the legitimate receiver and attacker (eavesdropper). 
In general, at least one Tx and one Rx electrode will be destroyed through an attack with a 250 to \SI{300}{\micro\meter} drill. 
As explained in Section~\ref{sec:keyGenEnclosure}, the PUF-response is analog, and a natural solution is to quantize it. A temperature change or a tampering attempt, as shown in Subsection~\ref{subsec:analysisPUF-response}, can change the value of the PUF-response. Thus, the changed value may fall into another quantization level other than the original one. Therefore, this effect can be modeled using a $q$-ary channel with non-zero crossover probabilities. We simulated the probability matrix for all possible symbol changes with 100,000 PUF-responses drawn from the PUF-distribution considering the changes in the PUF-response as modeled in Section~\ref{sec:keyGenEnclosure}.
We model the legitimate and attacker channel through $q$-ary channels for 8, 16, and 32 equiprobable intervals. In contrast to equidistant quantization, equiprobable intervals provide the required uniformly distributed input symbols at the polar encoder. 
A bias in the distribution of the input symbols can lead to additional leakage that has to be taken into consideration~\cite{CW19}. We observed that the channel for the legitimate receiver is not symmetric 
and therefore every $P(y_i\lvert c_j) = p_{i,j}$ can be different for $i,j \in \{0,1,\dots, q-1\}$.
We obtained the $p_{i,j}$ from simulating 100,000 PUF-responses for the legitimate and the attacker channel with the system model parameters described in Section~\ref{sec:keyGenEnclosure}. The simulation was done under consideration of the PUF-post-processing steps with and without analog helper data.

Since our PUF consists of 8 Tx pairs and 16 Rx electrodes, its response will contain $128$ values quantized into 128 $q$-ary symbols. We use a code-offset solution for error correction and, thus, need to construct a polar code of length $n = 128$. Note that the length of the polar code is independent of the value of the quantization level $q$. We construct the Polar Code through a Monte Carlo simulation~\cite{Arikan}. The core idea of the Monte Carlo code construction is to use uncoded transmission (encode $n$ symbols into a codeword of $n$ symbols) and to add noise to the codeword according to the channel model.  
Then, \gls{scd} decoding is performed over the received vector using the soft information of the channel model. 
The decoding is done as usual, except that after a hard decision, the decoder checks if the decision was correct or not, counts the occurrence of incorrect decisions, and --- if necessary --- changes the decision to the correct symbol. 
The idea behind this construction method is to estimate the probability that the decoding of a symbol $\hat{u}_i$ is wrong given correct previous hard decisions $(\hat{u}_1, \dots, \hat{u}_{i-1}) = (u_1, \dots, u_{i-1})$. If we ran this algorithm infinitely many times, the estimated probabilities would be the real ones, but a sufficiently good estimation can be achieved with finitely many runs. We perform this estimation method for both channels, namely the legitimate and the attacker one. We freeze the channels that are very bad (high probability of an erroneous decision) for both channels. We also transmit random symbols for the channels that are very good (probability of an erroneous decision $\approx 0$) and use all others for transmitting information symbols. We then calculate the entropy in bits for the attacker, by taking into account the estimations from the Monte Carlo construction. For every polarized channel symbol where information is transmitted, we check the distribution of a SC decoding. We checked that many outcomes were uniformly distributed among the extension field (ensuring maximum entropy), and some outcomes were leaning a bit more towards the correct symbol. However, the fact that we can ensure uniformly distributed input symbols (due to equiprobable quantization) does not allow the attacker to guess using the probability of each symbol (MAP decoding). An entropy of $s$ bits means that the attacker can at its best case brute-force $2^s$ options for the decoding process. For \gls{scl} decoding this means that the list size $L$ should be at least $L > 2^s$ large in order to have the correct codeword in the final list. Due to the code-offset construction, we add the PUF-response to a codeword of the polar code generated randomly and store the helper data.

\input{secs/sim/sim_results_8_intervals}
The results from the Monte Carlo code construction are listed in Table~\ref{tab:q8} for three different numbers of intervals $q$, with and without analog helper data $W'$.
Legitimate symbol channels with a probability for incorrect decoding below the threshold $d$ are selected to estimate the entropy of the attacker channel $H_\mathrm{att}$ in bits. 
Hence, this threshold is also a measure of the overall reliability of the legitimate channel.
The selection results in a number of symbols that are good for the legitimate channel $n_\mathrm{s}$.
This scheme can reliably reproduce $n_s$ symbols containing an entropy of $H_\mathrm{secret} = n_s \cdot \log_2 (q)$ bits.
The dimension of the code is $k = n_s$ in this wiretap construction.
$H_\mathrm{secret}$ corresponds to the bit length of the secret in conventional PUF key scenarios without wiretap secrecy leakage.
These bits can be directly used as a symmetrical encryption key or hashed to fit into a certain key length.

Of the reliable symbols $n_\mathrm{s}$, $n_\mathrm{f}$ symbols are also good for the attacker channel and have to be randomized. 
An attacker receives the remaining $n_s - n_f$ symbols exhibiting a high error rate from noisy to entirely random.
The complexity for an attacker is expressed through $H_\mathrm{att} = - \sum_{i}^{n_s} p_{s,i} \log_2 (p_{s,i})$, where $p_{s,i}$ denotes the symbol error rate after an attack.
$H_\mathrm{att}$ determines the physical layer security level of the wiretap PUF scheme.

As the parameter $d$ determines the number of symbols $n_s$, this parameter relates to the security level and enables a trade-off between security and reliability;
a stricter threshold $d$ selects only the most reliable bits, but also reduces the number of legitimate bits $n_s$ for the secret and the complexity for an attacker $H_\mathrm{att}$.

As the first Monte Carlo simulation only yielded per-symbol error rates, we performed another Monte Carlo simulation to obtain the \gls{fer}  for specific code construction parameters of Table~\ref{tab:q8} with the \gls{scd} and \gls{scl} decoder. The simulation results are listed in Table~\ref{tab:result_summary} for a temperature of $T=\ang{20}$. The \gls{fer} is the probability of the PUF secret being wrongly decoded.

The results show that we obtain a PUF-secret with up to $306$ bit length with 8 intervals, and $275$ bits for $32$ intervals, while reaching a \gls{fer} in the order of $10^{-6}$. The entropy of the attacker channel amounts to $100$ bits for 8 intervals. For 32 intervals, the brute force effort for the attacker is reduced to $2^{57}$, while still preserving $275$ bits of entropy for the PUF-secret.

\begin{table}[t!]
    \centering
    \begin{tabular}{|l|l|l|l|l|l|l|l|} \hline
        Decoder & $q$ & $W'$ & FER & $n_\mathrm{s}$ & $n_f$ & $H_\mathrm{att}$ & $H_\mathrm{secret}$ \\ \hline
        SCD & 8 & yes & $4.0 \times 10^{-6}$ & 102 & 22 & 100 & 306 \\ 
        SCL ($L=8$) & 8 & yes & $1.0 \times 10^{-6}$ & 102 & 22 & 100 & 306 \\
        SCD & 32 & no & $7.0 \times 10^{-6}$ & 55 & 11 & 57 & 275 \\ 
        SCL ($L=8$) & 32 & no & $3.3 \times 10^{-6}$ & 55 & 11 & 57 & 275 \\ \hline
    \end{tabular}
    \caption{Decoding results for T=\ang{20} using the $q$-ary Polar codes. 
    The number of correctly transmitted symbols is denoted as $n_s$.
    Of those $n_s$ symbols, $n_f$ symbols are readable in an attack scenario, while the remaining $n_s - n_f$ symbols exhibit a high error rate.
    With that, we estimate the secret length in bits as $H_\mathrm{secret}$ and the complexity of an attacker to $H_\mathrm{att}$.
    }
    \label{tab:result_summary}
\end{table}

The polar code design supports temperature changes in the range $\left[+\ang{5}, +\ang{35}\right]$, which is compatible with the operating temperatures of state-of-the-art network \glspl{hsm} \cite{thales-hsm, atalla-temperature}. 

Polar codes are praised for their low encoding and decoding complexity, especially for the binary case. The encoding and the \gls{scd} decoding can be performed in $\mathcal{O}(n \log n)$ time, while \gls{scl} decoding is performed in $\mathcal{O}(L \cdot n\log n)$~\cite{Tal2011,Balatsoukas-Stimming2015}. As for the $q-$ary polar codes, the check node and variable operations require $\mathcal{O}(q \log q)$ and $\mathcal{O}(q)$, respectively, instead of $\mathcal{O}(1)$ for the binary case. For a low number of intervals, the decoding complexity for $q$-ary codes reaches a low number of additional operations, which makes even $q$-ary polar codes suitable for implementation on a microcontroller. 

%% file: secs/wiretap_channel_pic.tex
\tikzstyle{block} = [rectangle, draw, text width=15em, text centered, minimum 		width=15em, minimum height=2em]
\tikzstyle{blockDiagonal} = [rectangle with diagonal fill, draw, text width=7em, text centered, minimum width=7em, minimum height=3em]
\tikzstyle{line} = [draw, -, thick]
\tikzstyle{arrow} = [draw, -latex, thick]
\tikzstyle{block} = [rectangle, draw, text centered, minimum height=2em]
\definecolor{lightergrey}{rgb}{0.83, 0.83, 0.83}

\begin{tikzpicture}
\draw[draw=black, fill=lightergrey, opacity=0.6] (-4.3,-3.85) rectangle (0.6,-0.4);
\node [draw=none, fill=none, xshift=-2.7cm, yshift=-0.85cm](keyEnrollment){\textcolor{darkgray}{\textbf{Key enrollment}}};
\draw[draw=black, fill=lightergrey, opacity=0.6] (3.3,-3.85) rectangle (9.05,-0.4);
\node [draw=none, fill=none, xshift=7.3cm, yshift=-0.85cm](keyReproduction){\textcolor{darkgray}{\textbf{Key reproduction}}};
\draw[dashed] (-0.15,-4.6)--(-0.15,-0.5);
\draw[dashed] (4.05,-4.6)--(4.05,-0.5);
\node [draw=none, fill=none](originPUF){};
\node [draw=none, below of=originPUF, node distance=1.3cm](quantization){};
\node [draw=none, fill=none, right of=quantization, node distance=2.0cm, yshift=-0.6cm, xshift=0.7cm](analogHelperData){$\hat{\epsilon}_a, p_2$};

\node [draw=none, fill=none, below of=quantization](quantizedPUF){};
\node [draw=none, fill=none, left of=quantizedPUF, below of=quantizedPUF, xshift=-0.0cm](codeword){$C$};
\node [block, left of=codeword](enc){enc};
\node [draw=none, fill=none, left of=enc, node distance = 1.4cm](trng){$R\equiv S$};
\node [draw=none, fill=none, below of=quantizedPUF](xor){};
\node [draw=none, fill=none, right of=xor, node distance=2.0cm](helperData){};
\node [draw=none, fill=none, above of=helperData, node distance=0.2cm, xshift=-0.5cm](helperData2){$\hat{\epsilon}_n, \hat{\epsilon}_t, p_1$};

\node [draw=none, fill=none, right of=helperData, right of=helperData](xor1){};
\node [draw=none, fill=none, right of=xor1, node distance=1.6cm](codeword2){$C+\hat{\epsilon}_{n}+\hat{\epsilon}_{t}$};
\node [draw=none, fill=none, above of=xor1](quantizedPUF2){};
\node [draw=none, above of=quantizedPUF2](quantization2){};
\node [draw=none, fill=none, right of=quantization2, node distance=1.6cm, yshift=-0.8cm](codeword3){$C+\hat{\epsilon}_{n}+\hat{\epsilon}_{t}+\hat{\epsilon}_{a}$};
\node [block, right of=codeword2, node distance=1.9cm](dec){dec};
\node [block, right of=codeword3, node distance=1.9cm](dec2){dec};
\node [draw=none, fill=none, right of=dec](secret){$S$};
\node [draw=none, fill=none, right of=dec2](secret2){$S'$};
\node [draw=none, fill=none, below of=helperData, node distance=1.1cm](bsc){noisy channels};
\node [draw=none, fill=none] at ([yshift=-0.1cm, xshift=2.3cm] xor) (aux){};

\path [arrow] (trng) -- (enc);
\path [arrow] (enc) -- (codeword);
\path [arrow] (codeword) -- (codeword2);
\path [arrow] (aux) |- (codeword3);
\path [arrow] (codeword2) -- (dec);
\path [arrow] (codeword3) -- (dec2);
\path [arrow] (dec) -- (secret);
\path [arrow] (dec2) -- (secret2);

\end{tikzpicture}

%% file: kernel.tex
	\begin{tikzpicture}[scale=1.5, every node/.style={scale=1}]
	\coordinate (L0) at (-1,0);
	\coordinate (L1) at (-1,-1);
	\coordinate (CH0) at (2,0);
	\coordinate (CH1) at (2,-1);
	\coordinate (R0) at (3,0);
	\coordinate (R1) at (3,-1);
	\coordinate (VN) at (0,-1);
	\coordinate (CN) at (0,0);
	
	\node[draw] (P0) at (CH0) {$P(y|c)$};
	\node[draw] (P1) at (CH1) {$P(y|c)$};
	\node[draw] (alpha) at (0,-0.5) {$\cdot \alpha$};

	\node[inner sep=0, outer sep=0] (OPLUS) at (CN) {\Large $\oplus$};
	\draw[] (L0)  -- (OPLUS) -- (P0) -- (R0);
	\draw[] (L1) -- (VN) -- (P1) -- (R1);
	\draw[] (VN) -- (alpha) -- (OPLUS);
	\node at (-1.2,0) {$u_1$};
	\node at (-1.2,-1) {$u_2$};
	\node at (0.7,0.2) {$c_1$};
	\node at (0.7,-0.8) {$c_2$};
	\node at (3.2,0) {$y_1$};
	\node at (3.2,-1) {$y_2$};
	\end{tikzpicture}

%% file: polar_code_figure.tex
\begin{tikzpicture}[scale=1.5, every node/.style={scale=1}]
	\draw[draw=black] (0,1) rectangle ++(2,3);
	\draw[draw=black] (0,4.4) rectangle ++(2,3);
	\node (u1) at (-0.6,7.2) {$u_1$}; \node (x1) at (4.6,7.35) {$c_1$};
	\node (u2) at (-0.6,6.6) {$u_2$}; \node (x2) at (4.6,6.75) {$c_2$};
	\node (uN/2) at (-0.6,4.7) {$u_{\frac{n}{2}}$}; \node (xN/2) at (4.6,4.95) {$c_{\frac{n}{2}}$};
	\path (u2) -- (uN/2) node [black, midway, sloped] {$\dots$};
	\path (4.6,6.35) -- (4.6,4.9) node [black, midway, sloped] {$\dots$}; 
	\node (uN/21) at (-0.6,3.8) {$u_{\frac{n}{2}+1}$}; \node (xN/21) at (4.6,4.05) {$c_{\frac{n}{2}+1}$};
	\node (uN/22) at (-0.6,3.2) {$u_{\frac{n}{2}+2}$}; \node (xN/22) at (4.6,3.45) {$c_{\frac{n}{2}+2}$};
	\node (uN) at (-0.6,1.2) {$u_{n}$}; \node (xN) at (4.6,1.35) {$c_{n}$};
	\path (uN/22) -- (uN) node [black, midway, sloped] {$\dots$};
	\path (4.6,3) -- (4.6,1.35) node [black, midway, sloped] {$\dots$};
	
	\node[inner sep=0, outer sep=0] (OPLUS1) at (4,7.2) {\Large $\oplus$};
	\node[inner sep=0, outer sep=0] (OPLUS2) at (3.7,6.6) {\Large $\oplus$};
	\node[inner sep=0, outer sep=0] (OPLUS3) at (2.8,4.7) {\Large $\oplus$};
	\node[draw] at (4, 5.5) (alpha1) {$\cdot \alpha$};
	\node[draw] at (3.7, 5) (alpha2) {$\cdot \alpha$};
	\node[draw] at (2.8, 2.5) (alpha3) {$\cdot \alpha$};
	\coordinate (VN1) at (4,3.8); \draw[fill] (VN1) circle (0.03);
	\coordinate (VN2) at (3.7,3.2); \draw[fill] (VN2) circle (0.03);
	\coordinate (VN3) at (2.8,1.2); \draw[fill] (VN3) circle (0.03);
	\path (OPLUS2) -- (OPLUS3) node [black, midway, sloped] {$\dots$};
	
	\coordinate (CH1) at (5.5,7.2); \coordinate (CH2) at (5.5,6.6); 
	\coordinate (CH3) at (5.5,4.7); \coordinate (CH4) at (5.5,3.8); 
	\coordinate (CH5) at (5.5,3.2); \coordinate (CH6) at (5.5,1.2);
	\node[draw] (P1) at (CH1) {$P(y|c)$}; \node[draw] (P2) at (CH2) {$P(y|c)$};
	\node[draw] (P3) at (CH3) {$P(y|c)$}; \node[draw] (P4) at (CH4) {$P(y|c)$};
	\node[draw] (P5) at (CH5) {$P(y|c)$}; \node[draw] (P6) at (CH6) {$P(y|c)$};
	\path (P2) -- (P3) node [black, midway, sloped] {$\dots$};
	\path (P5) -- (P6) node [black, midway, sloped] {$\dots$};
	
	\coordinate (M1) at (0,7.2); \coordinate (D1) at (2,7.2);
	\coordinate (M2) at (0,6.6); \coordinate (D2) at (2,6.6);
	\coordinate (Mn2) at (0,4.7); \coordinate (Dn2) at (2,4.7);
	\coordinate (M4) at (0,3.8); \coordinate (D4) at (2,3.8);
	\coordinate (M5) at (0,3.2); \coordinate (D5) at (2,3.2);
	\coordinate (M6) at (0,1.2); \coordinate (D6) at (2,1.2);
	\node (y1) at (6.6,7.2) {$y_1$}; \node (y2) at (6.6,6.6) {$y_2$};
	\node (yN/2) at (6.6,4.7) {$y_{n/2}$};
	\path (y2) -- (yN/2) node [black, midway, sloped] {$\dots$};
	\node (yN/21) at (6.6,3.8) {$y_{n/2+1}$}; \node (yN/22) at (6.6,3.2) {$y_{n/2+2}$};
	\node (yN) at (6.6,1.2) {$y_{n}$};
	\path (yN/22) -- (yN) node [black, midway, sloped] {$\dots$};
	\draw[] (u1) -- (M1); \draw[] (D1) -- (OPLUS1) -- (P1) -- (y1);
	\draw[] (u2) -- (M2); \draw[] (D2) -- (OPLUS2) -- (P2) --(y2);
	\draw[] (uN/2) -- (Mn2); \draw[] (Dn2) -- (OPLUS3)  -- (P3) --(yN/2);
	\draw[] (uN/21) -- (M4); \draw[] (D4) -- (VN1) -- (P4) -- (yN/21);
	\draw[] (uN/22) -- (M5); \draw[] (D5) -- (VN2) -- (P5) -- (yN/22);
	\draw[] (uN) -- (M6); \draw[] (D6) -- (VN3) -- (P6) -- (yN);
	\path (VN2) -- (VN3) node [black, midway, sloped] {$\dots$};
	\draw[] (VN1) -- (alpha1) -- (OPLUS1); \draw[] (VN2) -- (alpha2) -- (OPLUS2); \draw[] (VN3) -- (alpha3) -- (OPLUS3);
	
	\node[] (aaaa) at (1,2.5) {$\mathrm{F}_2(\alpha)^{\otimes \log\left(\frac{n}{2}\right)}$};
	\node[] (aaaa) at (1,5.8) {$\mathrm{F}_2(\alpha)^{\otimes \log\left(\frac{n}{2}\right)}$};
	
	\end{tikzpicture}

%% file: polar_decoder.tex
\begin{tikzpicture}[scale=1.5, every node/.style={scale=1}]
	\draw[draw=black] (0,1) rectangle ++(2,3);
	\draw[draw=black] (0,4.4) rectangle ++(2,3);
	\node (u1) at (-0.6,7.2) {$0$}; \node (x1) at (4.6,7.35) {$c_1$};
	\node (u2) at (-0.6,6.6) {$0$}; \node (x2) at (4.6,6.75) {$c_2$};
	\node (uN/2) at (-0.6,4.7) {$\hat{u}_{\frac{n}{2}}$}; \node (xN/2) at (4.6,4.95) {$c_{\frac{n}{2}}$};
	\path (u2) -- (uN/2) node [black, midway, sloped] {$\dots$};
	\path (4.6,6.35) -- (4.6,4.9) node [black, midway, sloped] {$\dots$}; 
	\node (uN/21) at (-0.6,3.8) {$\hat{u}_{\frac{n}{2}+1}$}; \node (xN/21) at (4.6,4.05) {$c_{\frac{n}{2}+1}$};
	\node (uN/22) at (-0.6,3.2) {$\hat{u}_{\frac{n}{2}+2}$}; \node (xN/22) at (4.6,3.45) {$c_{\frac{n}{2}+2}$};
	\node (uN) at (-0.6,1.2) {$\hat{u}_{n}$}; \node (xN) at (4.6,1.35) {$c_{n}$};
	\path (uN/22) -- (uN) node [black, midway, sloped] {$\dots$};
	\path (4.6,3) -- (4.6,1.35) node [black, midway, sloped] {$\dots$};
	
	\node[inner sep=0, outer sep=0] (OPLUS1) at (4,7.2) {\Large $\oplus$};
	\node[inner sep=0, outer sep=0] (OPLUS2) at (3.7,6.6) {\Large $\oplus$};
	\node[inner sep=0, outer sep=0] (OPLUS3) at (2.8,4.7) {\Large $\oplus$};
	\node at (4,7.5) (cnode1) {$\bar{c}_1$};
	\node at (3.7,6.9) (cnode2) {$\bar{c}_2$};
	\node at (2.8,5.0) (cnode3) {$\bar{c}_\frac{n}{2}$};
	\node[draw] at (4, 5.5) (alpha1) {$\cdot \alpha$};
	\node[draw] at (3.7, 5) (alpha2) {$\cdot \alpha$};
	\node[draw] at (2.8, 2.5) (alpha3) {$\cdot \alpha$};
	\coordinate (VN1) at (4,3.8); \draw[fill] (VN1) circle (0.03);
	\coordinate (VN2) at (3.7,3.2); \draw[fill] (VN2) circle (0.03);
	\coordinate (VN3) at (2.8,1.2); \draw[fill] (VN3) circle (0.03);
	\node at (4,3.5) (vnode1) {$v_1$};
	\node at (3.7,2.9) (vnode2) {$v_2$};
	\node at (2.8,0.9) (vnode3) {$v_\frac{n}{2}$};
	\path (OPLUS2) -- (OPLUS3) node [black, midway, sloped] {$\dots$};
	
	\coordinate (CH1) at (5.5,7.2); \coordinate (CH2) at (5.5,6.6); 
	\coordinate (CH3) at (5.5,4.7); \coordinate (CH4) at (5.5,3.8); 
	\coordinate (CH5) at (5.5,3.2); \coordinate (CH6) at (5.5,1.2);
	\node[draw] (P1) at (CH1) {$P(y|c)$}; \node[draw] (P2) at (CH2) {$P(y|c)$};
	\node[draw] (P3) at (CH3) {$P(y|c)$}; \node[draw] (P4) at (CH4) {$P(y|c)$};
	\node[draw] (P5) at (CH5) {$P(y|c)$}; \node[draw] (P6) at (CH6) {$P(y|c)$};
	\path (P2) -- (P3) node [black, midway, sloped] {$\dots$};
	\path (P5) -- (P6) node [black, midway, sloped] {$\dots$};
	
	\coordinate (M1) at (0,7.2); \coordinate (D1) at (2,7.2);
	\coordinate (M2) at (0,6.6); \coordinate (D2) at (2,6.6);
	\coordinate (Mn2) at (0,4.7); \coordinate (Dn2) at (2,4.7);
	\coordinate (M4) at (0,3.8); \coordinate (D4) at (2,3.8);
	\coordinate (M5) at (0,3.2); \coordinate (D5) at (2,3.2);
	\coordinate (M6) at (0,1.2); \coordinate (D6) at (2,1.2);
	\node (y1) at (6.6,7.2) {$y_1$}; \node (y2) at (6.6,6.6) {$y_2$};
	\node (yN/2) at (6.6,4.7) {$y_{n/2}$};
	\path (y2) -- (yN/2) node [black, midway, sloped] {$\dots$};
	\node (yN/21) at (6.6,3.8) {$y_{n/2+1}$}; \node (yN/22) at (6.6,3.2) {$y_{n/2+2}$};
	\node (yN) at (6.6,1.2) {$y_{n}$};
	\path (yN/22) -- (yN) node [black, midway, sloped] {$\dots$};
	\draw[] (u1) -- (M1); \draw[] (D1) -- (OPLUS1) -- (P1) -- (y1);
	\draw[] (u2) -- (M2); \draw[] (D2) -- (OPLUS2) -- (P2) --(y2);
	\draw[] (uN/2) -- (Mn2); \draw[] (Dn2) -- (OPLUS3)  -- (P3) --(yN/2);
	\draw[] (uN/21) -- (M4); \draw[] (D4) -- (VN1) -- (P4) -- (yN/21);
	\draw[] (uN/22) -- (M5); \draw[] (D5) -- (VN2) -- (P5) -- (yN/22);
	\draw[] (uN) -- (M6); \draw[] (D6) -- (VN3) -- (P6) -- (yN);
	\path (VN2) -- (VN3) node [black, midway, sloped] {$\dots$};
	\draw[] (VN1) -- (alpha1) -- (OPLUS1); \draw[] (VN2) -- (alpha2) -- (OPLUS2); \draw[] (VN3) -- (alpha3) -- (OPLUS3);
	
	\node[] (aaaa) at (1,2.5) {$\mathrm{F}_2(\alpha)^{\otimes \log\left(\frac{n}{2}\right)}$};
	\node[] (aaaa) at (1,5.8) {$\mathrm{F}_2(\alpha)^{\otimes \log\left(\frac{n}{2}\right)}$};
	
	\end{tikzpicture}

%% file: secs/sim/sim_results_8_intervals.tex
\begin{table}[t!]
    \centering
\begin{tabular}{|l|r r r r r|r r r r r|}
\hline
 &  \multicolumn{5}{c|}{Without $W'$}  &  \multicolumn{5}{c|}{With $W'$} \\ 
$q$  &  $d$ & $n_\mathrm{s}$ & $n_\mathrm{f}$ & $H_\mathrm{att}$ & $H_\mathrm{secret}$ &  $d$ & $n_\mathrm{s}$ & $n_\mathrm{f}$ & $H_\mathrm{att}$ &  $H_\mathrm{secret}$  \\ \hline
\multirow{8}*{8} & 
      0.0500 & 91 & 11 & 113.5 & 273 & 0.0500 & 123 & 22 & 163.0 & 369 \\
    & 0.0100 & 85 & 11 & 95.6  & 255 & 0.0100 & 121 & 22 & 157.9 & 363 \\
    & 0.0050 & 82 & 11 & 86.8  & 246 & 0.0050 & 120 & 22 & 154.9 & 360 \\
    & 0.0010 & 73 & 11 & 60.8  & 219 & 0.0010 & 119 & 22 & 151.9 & 357 \\
    & 0.0005 & 71 & 11 & 55.6  & 213 & 0.0005 & 117 & 22 & 145.9 & 351 \\
    & 0.0001 & 65 & 11 & 40.2  & 195 & 0.0001 & 112 & 22 & 130.9 & 336 \\
    & - &   &   &              &  & $10^{-5}$ & 106 & 22 & 112.0 & 318 \\  
    & $10^{-6}$ & 56 & 11 & 22.1 & 168 & $\mathbf{10^{-6}}$ & \textbf{102} & \textbf{22} & \textbf{100.3} & \textbf{306} \\ 
    & - &   &   &              &  & $<10^{-6}$ & 101 & 22 & 98.0 & 303 \\
\hline
\multirow{8}*{16} & 
      0.0500 & 76 & 9 & 110.0 & 304 & 0.0500 & 99 & 17 & 158.9 & 396 \\ 
    & 0.0100 & 70 & 9 & 87.5  & 280 & 0.0100 & 91 & 17 & 127.1 & 364 \\ 
    & 0.0050 & 67 & 9 & 76.3  & 268 & 0.0050 & 88 & 17 & 115.3 & 352 \\ 
    & 0.0010 & 61 & 9 & 55.7  & 244 & 0.0010 & 82 & 17 & 92    & 328 \\ 
    & 0.0005 & 58 & 9 & 47.8  & 232 & 0.0005 & 79 & 17 & 80.8  & 316 \\ 
    & 0.0001 & 52 & 9 & 33.1  & 208 & 0.0001 & 74 & 17 & 63.9  & 296 \\ 
    & - &   &   &             &  & $10^{-5}$ & 68 & 17 & 45.6  & 272 \\
    & $10^{-6}$ & 45 & 9 & 15 & 180 & $10^{-6}$ & 64 & 17 & 34.9 & 256 \\
     & - &   &   &            &  & $<10^{-6}$ & 63 & 17 & 32.0 & 252 \\
\hline
\multirow{8}*{32} & 
      0.0500 & 80 & 11 & 168.7 & 400 & 0.0500 & 86 & 15 & 181.0 & 430 \\ 
    & 0.0100 & 75 & 11 & 143.0 & 375 & 0.0100 & 78 & 15 & 141.4 & 390 \\ 
    & 0.0050 & 72 & 11 & 129.4 & 360 & 0.0050 & 76 & 15 & 131.7 & 380 \\ 
    & 0.0010 & 68 & 11 & 111.4 & 340 & 0.0010 & 73 & 15 & 116.9 & 365 \\ 
    & 0.0005 & 66 & 11 & 102.5 & 330 & 0.0005 & 72 & 15 & 112.4 & 360 \\ 
    & 0.0001 & 62 & 11 & 89.1  & 310 & 0.0001 & 69 & 15 & 98.9  & 345 \\
    & - &   &   &              &  & $10^{-5}$ & 62 & 15 & 74.0  & 310 \\
    & $\mathbf{10^{-6}}$ & \textbf{55} & \textbf{11} & \textbf{57.3} & \textbf{275} & $10^{-6}$ & 58 & 15 & 58.9 & 290 \\ 
    & - &   &   &              &  & $<10^{-6}$ & 56 & 15 & 49.9 & 280 \\
\hline 
\end{tabular}

\caption{The $q$-ary polar code construction results for 8, 16, and 32 intervals. The codeword consists of 128 symbols. Here, $d$ denotes the per-symbol error probability that indicates the error correction reliability. The number of reliable symbols after error correction is given as $n_s$, of which $n_f$ symbols are readable for an attacker. The attacker entropy $H_\mathrm{att}$ denotes the achievable security level from the remaining secret symbols in an attack scenario. We distinguish between decoding with and without analog helper data $W'$. 
}
\label{tab:q8}
\end{table}

%% file: secs/conclusion.tex
\section{Conclusion}
\label{sec:conclusion}

In this paper, we constructed a wiretap channel for the capacitive PUF-based enclosure from $q$-ary polar codes, and modeled the effects of attacks on the PUF-response.

First, we analyzed how temperature changes and drilling attacks affect the PUF-distribution based on real data obtained from measurements of the PUF-response. From this analysis, we derived a system model of the enclosure considering the impact of the PUF post-processing and different choices of quantization intervals. 

To construct the polar code for our Higher Order Alphabet PUF, we modeled the error behavior of the capacitive PUF-based enclosure through $q$-ary channels, and selected the best symbol channels for the legitimate wiretap channel, while minimizing the good symbol channels for the attacker. 
The wiretap code protects the information stored in these symbols, a code property that non-wiretap codes could not achieve.

With a Monte Carlo simulation for 8, 16, and 32 intervals and two different decoders, we demonstrated a physical layer security of 100 bits, while preserving 306 bits of entropy for the PUF-secret.

%% file: main.bbl
\newcommand{\etalchar}[1]{$^{#1}$}
\begin{thebibliography}{HWRL{\etalchar{+}}13}

\bibitem[{Adv}12]{FIPS-IBM}
{Advanced Cryptographic Hardware Development IBM Poughkeepsie and IBM
  Research}.
\newblock {\em {IBM 4765 cryptographic coprocessor security modulesecurity
  policy}}.
\newblock Zürich, December 2012.

\bibitem[Ari09]{Arikan}
Erdal Arikan.
\newblock Channel polarization: A method for constructing capacity-achieving
  codes for symmetric binary-input memoryless channels.
\newblock {\em IEEE Transactions on information Theory}, 55(7):3051--3073,
  2009.

\bibitem[BB19]{NIST-SP800-57}
Elaine Barker and William~C. Barker.
\newblock {\em {Recommendation for Key Management: Part 2 -- Rev. 1 -- Best
  Practices for Key Management Organizations}}.
\newblock National Institute of Standards and Technology, Gaithersburg, MD, May
  2019.

\bibitem[BCC{\etalchar{+}}12]{Briais2012}
S{\'e}bastien Briais, St{\'e}phane Caron, Jean-Michel Cioranesco, Jean-Luc
  Danger, Sylvain Guilley, Jacques-Henri Jourdan, Arthur Milchior, David
  Naccache, and Thibault Porteboeuf.
\newblock 3d hardware canaries.
\newblock In Emmanuel Prouff and Patrick Schaumont, editors, {\em Cryptographic
  Hardware and Embedded Systems -- CHES 2012}, pages 1--22, Berlin, Heidelberg,
  2012. Springer Berlin Heidelberg.

\bibitem[BGS{\etalchar{+}}08]{BGS+08}
Christoph B{\"o}sch, Jorge Guajardo, Ahmad-Reza Sadeghi, Jamshid Shokrollahi,
  and Pim Tuyls.
\newblock Efficient helper data key extractor on fpgas.
\newblock In {\em IACR CHES}, pages 181--197. Springer Berlin Heidelberg, 2008.

\bibitem[BSPB15]{Balatsoukas-Stimming2015}
Alexios Balatsoukas-Stimming, Mani~Bastani Parizi, and Andreas Burg.
\newblock Llr-based successive cancellation list decoding of polar codes.
\newblock {\em IEEE Transactions on Signal Processing}, 63(19):5165--5179,
  2015.

\bibitem[BY19]{BY19}
Yonghong Bai and Zhiyuan Yan.
\newblock A secure and robust key generation method using physical unclonable
  functions and polar codes.
\newblock In {\em 2019 IEEE International Workshop on Signal Processing Systems
  (SiPS)}, pages 254--259, 2019.

\bibitem[BY21]{Bai2021}
Yonghong Bai and Zhiyuan Yan.
\newblock A secure and robust puf-based key generation with wiretap polar coset
  codes.
\newblock {\em Journal of Electronic Testing}, 37, 06 2021.

\bibitem[CIW{\etalchar{+}}17]{CIW+17}
Bin Chen, Tanya Ignatenko, Frans M.~J. Willems, Roel Maes, Erik van~der Sluis,
  and Georgios Selimis.
\newblock A robust sram-puf key generation scheme based on polar codes.
\newblock In {\em GLOBECOM 2017 - 2017 IEEE Global Communications Conference},
  pages 1--6, 2017.

\bibitem[CK06]{Csiszar}
I.~Csiszar and J.~Korner.
\newblock Broadcast channels with confidential messages.
\newblock {\em IEEE Trans. Inf. Theor.}, 24(3):339–348, sep 2006.

\bibitem[CRR98]{CRR98}
R.~Chandramouli, N.~Ranganathan, and S.J. Ramadoss.
\newblock Adaptive quantization and fast error-resilient entropy coding for
  image transmission.
\newblock {\em IEEE Transactions on Circuits and Systems for Video Technology},
  8(4):411--421, 1998.

\bibitem[CS16]{chauhan2016fuzzy}
Sonam Chauhan and Ajay Sharma.
\newblock Fuzzy commitment scheme based on reed solomon codes.
\newblock In {\em Proceedings of the 9th International Conference on Security
  of Information and Networks}, pages 96--99, 2016.

\bibitem[CW19]{CW19}
Bin Chen and Frans M.~J. Willems.
\newblock Secret key generation over biased physical unclonable functions with
  polar codes.
\newblock {\em IEEE Internet of Things Journal}, 6(1):435--445, 2019.

\bibitem[DORS08]{Dodis2008}
Yevgeniy Dodis, Rafail Ostrovsky, Leonid Reyzin, and Adam Smith.
\newblock Fuzzy extractors: How to generate strong keys from biometrics and
  other noisy data.
\newblock {\em SIAM journal on computing}, 38(1):97--139, 2008.

\bibitem[DRS04]{fuzzyExtractor}
Yevgeniy Dodis, Leonid Reyzin, and Adam Smith.
\newblock Fuzzy extractors: How to generate strong keys from biometrics and
  other noisy data.
\newblock In {\em IACR Eurocrypt}, pages 523--540. Springer, 2004.

\bibitem[EFK{\etalchar{+}}12]{EFK+12}
Thomas Esbach, Walter Fumy, Olga Kulikovska, Dominik Merli, Dieter Schuster,
  and Frederic Stumpf.
\newblock A new security architecture for smartcards utilizing pufs.
\newblock In {\em Information Security Solutions Europe (ISSE 2012)}. Vieweg
  Verlag, 2012.

\bibitem[{Fed}08]{Fed08}
{Federal Office for Information Security}.
\newblock {Common Criteria Protection Profile -- Cryptographic Modules,
  Security Level “Enhanced” (BSI-CC-PP-0045)}, 2008.

\bibitem[FIU{\etalchar{+}}18]{ASIC}
E.~{Ferres}, V.~{Immler}, A.~{Utz}, A.~{Stanitzki}, R.~{Lerch}, and
  R.~{Kokozinski}.
\newblock Capacitive multi-channel security sensor ic for tamper-resistant
  enclosures.
\newblock In {\em IEEE SENSORS}, pages 1--4, 2018.

\bibitem[GN98]{GN98}
R.M. Gray and D.L. Neuhoff.
\newblock Quantization.
\newblock {\em IEEE Transactions on Information Theory}, 44(6):2325--2383,
  1998.

\bibitem[GOFK21]{FORTRESS}
Kathrin Garb, Johannes Obermaier, Elischa Ferres, and Martin König.
\newblock Fortress: Fortified tamper-resistant envelope with embedded security
  sensor.
\newblock In {\em 18th Annual International Conference on Privacy, Security and
  Trust (PST2021)}, 2021.

\bibitem[Gro]{thales-hsm}
Thales Group.
\newblock High assurance hardware security modules.
\newblock
  \url{https://cpl.thalesgroup.com/encryption/hardware-security-modules/network-hsms}.
\newblock Online, accessed on 16 January 2022.

\bibitem[GSHO21]{GSHO21}
Kathrin Garb, Marc Schink, Matthias Hiller, and Johannes Obermaier.
\newblock Attacks and countermeasures for capacitive puf-based security
  enclosures.
\newblock In {\em 2021 IEEE Physical Assurance and Inspection of Electronics
  (PAINE)}, pages 1--8, 2021.

\bibitem[{Hew}]{atalla-temperature}
{Hewlett-Packard Company}.
\newblock {Hewlett-Packard --- Atalla Security Products Ax160 PCI HSM Security
  Policy}.
\newblock
  \url{https://www.pcisecuritystandards.org/ptsdocs/HP%20Atalla%20Ax160%20PCI%20HSM%20Security%20Policy%201_1.pdf}.
\newblock Online, accessed 14 November 2021.

\bibitem[HKS20]{HKS20}
M.~Hiller, Ludwig Kurzinger, and G.~Sigl.
\newblock Review of error correction for pufs and evaluation on
  state-of-the-art fpgas.
\newblock {\em Journal of Cryptographic Engineering}, 10:229--247, 2020.

\bibitem[HMSS12]{Hiller2012}
Matthias Hiller, Dominik Merli, Frederic Stumpf, and Georg Sigl.
\newblock Complementary ibs: Application specific error correction for pufs.
\newblock In {\em IEEE HOST}, pages 1--6. IEEE, 2012.

\bibitem[H{\"O}17]{hiller2017hiding}
Matthias Hiller and Aysun~Gurur {\"O}nalan.
\newblock Hiding secrecy leakage in leaky helper data.
\newblock In {\em International Conference on Cryptographic Hardware and
  Embedded Systems}, pages 601--619. Springer, 2017.

\bibitem[HOSB16]{HOSB16}
Matthias Hiller, Aysun~Gurur \"{O}nalan, Georg Sigl, and Martin Bossert.
\newblock Online reliability testing for puf key derivation.
\newblock In {\em TrustED}, page 15–22, New York, NY, USA, 2016. Association
  for Computing Machinery.

\bibitem[HWRL{\etalchar{+}}13]{Hiller2013a}
Matthias Hiller, Michael Weiner, Leandro Rodrigues~Lima, Maximilian Birkner,
  and Georg Sigl.
\newblock Breaking through fixed puf block limitations with differential
  sequence coding and convolutional codes.
\newblock In {\em TrustED}, pages 43--54, 2013.

\bibitem[HYP15]{Hiller2015a}
Matthias Hiller, Meng-Day Yu, and Michael Pehl.
\newblock Systematic low leakage coding for physical unclonable functions.
\newblock In {\em ASIA CCS}, pages 155--166, 2015.

\bibitem[IHKS16]{IHKS16}
Vincent Immler, Maxim Hennig, Ludwig K{\"u}rzinger, and Georg Sigl.
\newblock Practical aspects of quantization and tamper-sensitivity for
  physically obfuscated keys.
\newblock In {\em ACM CS2}, pages 13--18, 2016.

\bibitem[IHL{\etalchar{+}}19]{IHL+19}
Vincent Immler, Matthias Hiller, Qinzhi Liu, Andreas Lenz, and Antonia
  Wachter-Zeh.
\newblock Variable-length bit mapping and error-correcting codes for
  higher-order alphabet pufs—extended version.
\newblock {\em Journal of Hardware and Systems Security}, 3(1):78--93, 2019.

\bibitem[IMJFC13]{Gore-envelope}
Phil Isaacs, Thomas Morris~Jr, Michael~J Fisher, and Keith Cuthbert.
\newblock Tamper proof, tamper evident encryption technology.
\newblock In {\em Pan Pacific Symposium}, 2013.

\bibitem[IOK{\etalchar{+}}18]{B-TREPID}
Vincent Immler, Johannes Obermaier, Martin K{\"o}nig, Matthias Hiller, and
  Georg Sigl.
\newblock B-trepid: batteryless tamper-resistant envelope with a puf and
  integrity detection.
\newblock In {\em IEEE HOST}, pages 49--56. IEEE, 2018.

\bibitem[ION{\etalchar{+}}19]{COVER}
Vincent Immler, Johannes Obermaier, Kuan~Kuan Ng, Fei~Xiang Ke, JinYu Lee,
  Yak~Peng Lim, Wei~Koon Oh, Keng~Hoong Wee, and Georg Sigl.
\newblock Secure physical enclosures from covers with tamper-resistance.
\newblock {\em IACR CHES}, pages 51--96, 2019.

\bibitem[ISO12]{ISO-19790}
ISO, Geneva, Switzerland.
\newblock {\em {ISO/IEC 19790:2012 Information technology -- Security
  techniques -- Test requirements for cryptographic modules}}, August 2012.

\bibitem[ISO17]{ISO-24759}
ISO, Geneva, Switzerland.
\newblock {\em {ISO/IEC 24759:2017 Information technology -- Security
  techniques -- Test requirements for cryptographic modules}}, March 2017.

\bibitem[IU19]{LMC}
Vincent Immler and Karthik Uppund.
\newblock New insights to key derivation for tamper-evident physical unclonable
  functions.
\newblock {\em IACR CHES}, pages 30--65, 2019.

\bibitem[JL12]{originalLMC}
Myeongwoon Jeon and Jungwoo Lee.
\newblock On codes correcting bidirectional limited-magnitude errors for flash
  memories.
\newblock In {\em IEEE ISITA}, pages 96--100, 2012.

\bibitem[JW99]{fuzzyCommitment}
Ari Juels and Martin Wattenberg.
\newblock A fuzzy commitment scheme.
\newblock In {\em Proceedings of the 6th ACM Conference on Computer and
  Communications Security}, CCS '99, page 28–36, New York, NY, USA, 1999.
  Association for Computing Machinery.

\bibitem[Mae12]{MaesPhDThesis}
Roel Maes.
\newblock {\em Physically unclonable functions: Constructions, properties and
  applications}.
\newblock PhD thesis, Katholieke Universiteit Leuven, Belgium, 2012.

\bibitem[MB17]{MB17}
Sven Mueelich and Martin Bossert.
\newblock A new error correction scheme for physical unclonable functions.
\newblock In {\em SCC 2017; 11th International ITG Conference on Systems,
  Communications and Coding}, pages 1--6, 2017.

\bibitem[MHK{\etalchar{+}}19]{MHK+19}
H.~{Mandry}, A.~{Herkle}, L.~{Kürzinger}, S.~{Müelich}, J.~{Becker}, R.~F.~H.
  {Fischer}, and M.~{Ortmanns}.
\newblock Modular puf coding chain with high-speed reed-muller decoder.
\newblock In {\em IEEE ISCAS}, pages 1--5, 2019.

\bibitem[MLSW16]{MLSW16}
Roel Maes, Vincent Leest, Erik Sluis, and Frans Willems.
\newblock Secure key generation from biased pufs: extended version.
\newblock {\em Journal of Cryptographic Engineering}, 6, 06 2016.

\bibitem[MPB18]{MPB18}
Sven Müelich, Sven Puchinger, and Martin Bossert.
\newblock Using convolutional codes for key extraction in sram physical
  unclonable functions, 2018.

\bibitem[MTV09a]{MTV09a}
R.~{Maes}, P.~{Tuyls}, and I.~{Verbauwhede}.
\newblock A soft decision helper data algorithm for sram pufs.
\newblock In {\em IEEE ISIT}, pages 2101--2105, 2009.

\bibitem[MTV09b]{MTV09b}
Roel Maes, Pim Tuyls, and Ingrid Verbauwhede.
\newblock Low-overhead implementation of a soft decision helper data algorithm
  for sram pufs.
\newblock In {\em International workshop on cryptographic hardware and embedded
  systems}, pages 332--347. Springer, 2009.

\bibitem[MV11]{polar_wiretap}
Hessam Mahdavifar and Alexander Vardy.
\newblock Achieving the secrecy capacity of wiretap channels using polar codes.
\newblock {\em IEEE Transactions on Information Theory}, 57(10):6428--6443,
  2011.

\bibitem[MVHV12]{MVHV12}
Roel Maes, Anthony Van~Herrewege, and Ingrid Verbauwhede.
\newblock Pufky: A fully functional puf-based cryptographic key generator.
\newblock In Emmanuel Prouff and Patrick Schaumont, editors, {\em IACR CHES},
  pages 302--319, Berlin, Heidelberg, 2012. Springer Berlin Heidelberg.

\bibitem[Obe19]{ObermaierPhD}
Johannes Obermaier.
\newblock {\em Breaking and Restoring Embedded System Security}.
\newblock PhD thesis, Technische Universit{\"a}t M{\"u}nchen, 2019.

\bibitem[OHHS18]{EKMS}
Johannes Obermaier, Florian Hauschild, Matthias Hiller, and Georg Sigl.
\newblock An embedded key management system for puf-based security enclosures.
\newblock In {\em MECO}, pages 1--6. IEEE, 2018.

\bibitem[OI18]{OI18}
Johannes Obermaier and Vincent Immler.
\newblock The past, present, and future of physical security enclosures: from
  battery-backed monitoring to puf-based inherent security and beyond.
\newblock {\em Journal of Hardware and Systems Security}, 2(4):289--296, 2018.

\bibitem[OIHS18]{measurementCircuit}
Johannes Obermaier, Vincent Immler, Matthias Hiller, and Georg Sigl.
\newblock A measurement system for capacitive puf-based security enclosures.
\newblock In {\em DAC}, New York, NY, USA, 2018. Association for Computing
  Machinery.

\bibitem[OW85]{WynerII}
L.~H. Ozarow and A.~D. Wyner.
\newblock Wire-tap channel ii.
\newblock In Thomas Beth, Norbert Cot, and Ingemar Ingemarsson, editors, {\em
  Advances in Cryptology}, pages 33--50, Berlin, Heidelberg, 1985. Springer
  Berlin Heidelberg.

\bibitem[PB12]{polar_codes_q_ary_capacity}
Woomyoung Park and Alexander Barg.
\newblock Polar codes for q-ary channels, $q =2^{r}$.
\newblock In {\em 2012 IEEE International Symposium on Information Theory
  Proceedings}, pages 2142--2146, 2012.

\bibitem[PMB{\etalchar{+}}15]{PMB+15}
S.~{Puchinger}, S.~{Mueelich}, M.~{Bossert}, M.~{Hiller}, and G.~{Sigl}.
\newblock On error correction for physical unclonable functions.
\newblock In {\em ITG SCC}, pages 1--6, 2015.

\bibitem[SFIC14]{OpticalPUF}
Merrielle Spain, Benjamin Fuller, Kyle Ingols, and Robert Cunningham.
\newblock Robust keys from physical unclonable functions.
\newblock In {\em IEEE HOST}, pages 88--92. IEEE, 2014.

\bibitem[Sto03]{StoltePhD}
Norbert Stolte.
\newblock {\em Recursive Codes with the Plotkin-Construction and Their
  Decoding}.
\newblock PhD thesis, Technische Universit{\"a}t Darmstadt, 2003.

\bibitem[STZP21]{Staat2021}
Paul Staat, Johannes Tobisch, Christian~T. Zenger, and Christof Paar.
\newblock Anti-tamper radio: System-level tamper detection for computing
  systems.
\newblock {\em CoRR}, abs/2112.09014, 2021.

\bibitem[TS{\v{S}}{\etalchar{+}}06]{CoatingPUF}
Pim Tuyls, Geert-Jan Schrijen, Boris {\v{S}}kori{\'{c}}, Jan van Geloven, Nynke
  Verhaegh, and Rob Wolters.
\newblock Read-proof hardware from protective coatings.
\newblock In {\em IACR CHES}, pages 369--383. Springer Berlin Heidelberg, 2006.

\bibitem[TV11]{Tal2011}
Ido Tal and Alexander Vardy.
\newblock List decoding of polar codes.
\newblock In {\em 2011 IEEE International Symposium on Information Theory
  Proceedings}, pages 1--5, 2011.

\bibitem[TZP20]{TZP20}
Johannes Tobisch, Christian Zenger, and Christof Paar.
\newblock Electromagnetic enclosure {PUF} for tamper proofing commodity
  hardware and other applications.
\newblock In {\em TRUDEVICE}, 2020.

\bibitem[VNK{\etalchar{+}}15]{OpticalPUFSystem}
M.~{Vai}, B.~{Nahill}, J.~{Kramer}, M.~{Geis}, D.~{Utin}, D.~{Whelihan}, and
  R.~{Khazan}.
\newblock Secure architecture for embedded systems.
\newblock In {\em IEEE HPEC}, pages 1--5, Sep. 2015.

\bibitem[W.L]{Gore-cover}
W.L. GORE \& Associates, Inc., Electronic Products Division, Newark.
\newblock {\em {GORE$^\mathrm{TM}$ Tamper Respondent Surface Enclosure}}.

\bibitem[Wyn75]{WynerI}
A.~D. Wyner.
\newblock The wire-tap channel.
\newblock {\em The Bell System Technical Journal}, 54(8):1355--1387, 1975.

\bibitem[YD10]{IBS}
Meng-Day Yu and Srinivas Devadas.
\newblock Secure and robust error correction for physical unclonable functions.
\newblock {\em IEEE Design \& Test of Computers}, 27(1):48--65, 2010.

\bibitem[YHD15]{YHD15}
Meng-Day Yu, Matthias Hiller, and Srinivas Devadas.
\newblock Maximum-likelihood decoding of device-specific multi-bit symbols for
  reliable key generation.
\newblock In {\em 2015 IEEE International Symposium on Hardware Oriented
  Security and Trust (HOST)}, pages 38--43. IEEE, 2015.

\bibitem[YS18]{Yuan_Steiner}
Peihong Yuan and Fabian Steiner.
\newblock Construction and decoding algorithms for polar codes based on 2 × 2
  non-binary kernels.
\newblock In {\em 2018 IEEE 10th International Symposium on Turbo Codes
  Iterative Information Processing (ISTC)}, pages 1--5, 2018.

\bibitem[ZHW{\etalchar{+}}21]{ZHW+21}
Yin Zhang, Zhangqing He, Meilin Wan, Jiuyang Liu, Haoshang Gu, and Xuecheng
  Zou.
\newblock A {SC} {PUF} standard cell used for key generation and
  anti-invasive-attack protection.
\newblock {\em IEEE Transactions on Information Forensics and Security}, pages
  1--1, 2021.

\end{thebibliography}
